# Geographic Variation in Stack Overflow Code Quality: Evidence from a Cross-Regional Study of Coding Practices

Elijah Zolduoarrati*, Sherlock A. Licorish, Nigel Stanger
Department of Information Science
University of Otago
Dunedin, New Zealand
[elijah.zolduoarrati, sherlock.licorish, nigel.stanger]@otago.ac.nz

## ABSTRACT

Developers routinely integrate Stack Overflow code snippets into their codebases, making the platform ubiquitous. However, the quality of snippets embedded in users' answers remain elusive. Existing evaluations of code quality tend to be language-specific, necessitating holistic investigations that span multiple languages. Moreover, literature has found that contribution patterns vary depending on geographical locales, creating an unexplained rift between code quality, user location, and latent contextual regional factors. This study evaluates the quality of SQL, JavaScript, Python, Ruby, and Java snippets across reliability, readability, performance, and security dimensions. We benchmark findings across states in the US, assessing how different diversity indicators correlate against code quality violations. We conducted a series of inductive content analyses that qualitatively supplement prior quantitative findings. Outcomes show that readability violations are the most prevalent across all languages, followed by reliability, performance, and security. Readability violations largely comprise improper whitespace, while reliability issues mainly involve program flow alterations. Performance violations encompass resource allocation while security vulnerabilities include unvalidated inputs that risk injection attacks. Tech hubs like California show moderate reliability and readability violations, while tech-sparse areas like Maine exhibit fewer reliability and performance violations. Regions with equitable wealth distribution and higher per capita income tend to have better code quality, possibly due to increased investment in education and access to technology. Disparities in coding practices were also surmised to stem from a combination of gender-inherent traits, diversified tech workforce, emerging non-tech industries, and the prevalence of startup culture within these regions.

## 1 INTRODUCTION

In the digital age, the rapid evolution of technology has ushered in an era of unprecedented collaboration and information sharing [1]. Community Question Answering (CQA) platforms, exemplified by popular websites like Stack Overflow[1] and Quora[2], represent a remarkable outcome of this digital revolution [2]. These platforms have revolutionised information dissemination by enabling users to pose questions in natural language and receive tailored responses [2]. Stack Overflow, in particular, stands at the digital epicentre, where developers gather to share insights, seek guidance, and collaborate with peers regarding the technical challenges they face [3, 4]. The hallmark of this platform is its emphasis on high-quality content, including accurate, well-structured posts that provide robust solutions. Posts on Stack Overflow provide developers with more than just answers; they offer well-crafted solutions that address the root causes of issues, provide context, and illuminate underlying computer science-related concepts. Therefore, it is paramount that the information presented on Stack Overflow is of high quality, so that developers facing similar challenges are empowered to make informed decisions and build more and more robust solutions in the future.

Owing to its importance, quality evaluations on Stack Overflow have garnered major popularity in recent literature. However, scholarly works have primarily focussed on semantic (e.g., readability or clarity) and community-related factors (e.g., popularity or content curation activity of users) [5, 6, 7]. Studies focussed on semantics have oftentimes side-lined the facet of code quality, which is no less important given that developers frequently use Stack Overflow code snippets in their projects [8].

* Corresponding author
[1] https://stackoverflow.com
[2] https://www.quora.com

Existing evaluations of code quality tend to be language-specific instead of encompassing the broader picture. For instance, JavaScript, Java, and Python code snippets were investigated separately in three different research endeavours [9, 10, 11]. Moreover, previous work [12] has hinted that diversity complications in the tech field also propagate to Stack Overflow. For instance, it was shown that diversity-related problems range across human-inherent features (e.g., age and gender), across socio-economic constructs, technological accessibility, and workforce concentration [12]. While subsequent literature has tried to bridge the gap between contribution and diversity [13], the unexplained rift between code quality and the construct of diversity remains elusive. Thus, the need for a research agenda arises aimed at integrating these different facets under a unified framework.

This work aims to comprehensively investigate code quality in Stack Overflow by assessing multiple quality dimensions, namely reliability, readability, performance, and security. These dimensions were drawn from recent literature, acknowledging the absence of a singular, universally applicable definition for the construct of *quality* [11]. In addition, we progress the agenda of exploring diversity and how this affects SE-related practices by investigating regional variations in coding practices, aiming to shed light on how different regions vary with regard to producing reliable, readable, performant, and secure code. Our study builds upon the work of Meldrum et al. [11], whose approach we largely adopt in this investigation, particularly in terms of how *code quality* is defined. We leverage the Stack Overflow Database obtained from the Stack Exchange Data Dump, which has been used previously [7, 14]. Our study focusses exclusively on answers, aligning with user behaviour, which predominantly involves seeking answers and copying code solutions [15, 16]. Further, we narrow our investigation to examine code snippets originating from the United States of America (US) for several reasons[3]. Firstly, Stack Overflow's origin in 2008 was in the US before its global expansion [17]. Secondly, as the largest English-speaking nation [18], the US serves as a suitable setting for studying software development, as English is the *lingua franca* within this field [19]. Third, the US has been consistently positioned at the forefront of technological innovation [20, 21], providing a diverse pool of developers and access to cutting-edge resources. Finally, the prevalence of US users on Stack Overflow is among the highest in comparison to other countries, rendering it a representative platform for exploring the dynamics of contributors in CQA settings [12]. While Stack Overflow data is primarily user-generated (e.g., posts, comments, profile bios), prior research has demonstrated its usefulness as a proxy for exploring SE-related practices [22, 23]. Moreover, previous work had subjected the data to rigorous checks [12] and thus the data quality employed in this work is assured. SQL, JavaScript, Python, Ruby, and Java were selected for coverage in our study due to their widespread adoption among all US users (refer to Section 3.3). All five languages have also been studied extensively in the code quality niche [24, 25], yet findings from prior works remain compartmentalised.

To focus our study and achieve the planned objective, we have formulated the following research questions:

**RQ1**: What is the quality of code snippets across different programming languages on Stack Overflow?
**RQ2**: How do coding practices among US contributors differ across states and cities in the United States?
**RQ3**: How do diversity indicators of US regions affect code quality?
**RQ4**: What themes are prevalent in code quality violations, and how do these themes differ across US regions?

RQ1 provides insights into the quality of Stack Overflow code snippets, frequently used by developers, with respect to four quality dimensions: reliability, readability, performance, and security. The motivation behind RQ1 is to assess the platform's effectiveness in providing trustworthy solutions, as well as to ascertain whether the given solutions are comprehensible, performant, and are free from security vulnerabilities. RQ2 presents a granular analysis across all states and cities throughout the US to explore disparities in reliable, readable, performant, and secure coding practices. This will shed light on the causative factors behind such observed shifts, in turn identifying why these variations occur due to inherent contextual and cultural factors within each region. RQ3 focusses on identifying regional factors that may influence code quality within Stack Overflow snippets. We aim to understand how latent characteristics of a region, such as socioeconomic factors, access to infrastructure, or tech and R&D density may be able to impact the reliability, readability, performance, and security of the code snippets contributed by developers living in those regions. RQ4 delves into the nature of code quality violations, enabling the qualitative identification of any recurring themes across the four key dimensions of code quality. For example, within the dimension of readability, issues related to improper whitespace formatting may be more prevalent, as opposed to less widespread problems like function annotation or poorly named variables. We surmise that a deeper qualitative exploration will enrich the community's understanding into which specific types of code quality violations are prevalent.

[3] Our data covers 50 states and the District of Columbia. Our geocoding API (refer to Section 3.2) treated the US territories, namely American Samoa, Guam, Northern Mariana Islands, Puerto Rico, US Minor Outlying Islands, and US Virgin Islands, as distinct entities, but the US Census Bureau excludes these territories from the main 50 states.

Addressing these research questions offers contributions for both theory and practice. Our study contributes to existing literature by bridging a theoretical gap wherein multiple programming languages are assessed under the same quality dimensions. We expand this line of inquiry by taking into account census diversity indicators, thereby bridging the gap between technology and socioeconomics. From a practitioner standpoint, our insights promote reliable, readable, performant, and secure coding practices, empowering developers to adopt these enhancements into their daily development processes. As developers adopt higher-quality coding practices, the overall codebase of projects can improve, leading to better-performing, more maintainable, and more secure repositories. A qualitative examination of common themes within each code quality dimension is also presented, which may empower practitioners to proactively avoid common pitfalls in their own software development practices. Finally, while generative language models (e.g., ChatGPT or Gemini) have been linked to a decrease in overall Stack Overflow participation, their solutions are mostly error-prone [26]. Conversely, Stack Overflow offers solutions that are vetted by the community and specifically tailored to address niche problems, and thus its unique value stems from the quality of the shared code snippets. Our study thus remains valuable for developers, enabling them to write and identify high-quality code even in the age of generative language models.

The remainder of this manuscript is organised as follows. Section 2 provides the theoretical underpinnings of our study. Section 3 elucidates the methodologies employed for data collection, processing, and analysis. Section 4 presents the findings of the study with respect to each research question, before we discuss our results in Section 5. In Section 6, we discuss the limitations of the study, and we conclude with final remarks and suggestions for future research in Section 7. Additionally, a replication package is provided for those interested in further examining our research methodology and performing follow up analyses [27][4].

# 2 BACKGROUND

## 2.1 Diversity in Software Engineering

The inherent socio-technical nature of software engineering (SE) makes it prone to the accumulation of social debt, particularly related to diversity within the field [28]. However, the relationship between developer background diversity and project outcomes remains elusive. On the one hand, diversity in factors like age, gender, culture, and education can foster synergy, creativity, and a broader range of perspectives when tackling problems [29, 30]. On the other, they can also lead to increased potential for conflict [31]. Language barriers are also a threat, where the dominance of English as the *lingua franca* in software development creates a barrier for non-native speakers, hindering their ability to keep pace with advances in the field [32]. Unequal access to infrastructure and communication services, also known as the *digital divide*, further exacerbates the existing challenges of diversity within SE-related practices. This disparity permeates to all layers of society across individuals, households, businesses, and entire geographical regions, and has become even more pronounced since the emergence of COVID-19 [33]. These factors result in a complex interplay that impedes diversity in SE [12]. External challenges, such as limited access to infrastructure and economic disparities, act as barriers to participation. Internally, unconscious biases and exclusionary practices can stem from inherent human characteristics like educational background, age, gender, and cultural constructs [12]. In fact, seemingly technical challenges in SE-related processes such as insufficient documentation and software defects are often intertwined with latent, underlying human aspects [34]. To this end, a comprehensive tertiary review regarding what human factors influence SE processes has emphasised the need for tailored approaches to bridge diversity gaps, rather than resorting to one-size-fits-all solutions [34]. While research has extensively explored individual human aspects like gender, race, and culture [35, 36, 37], external factors such as demographics and socioeconomic background remain understudied [34].

[4] https://zenodo.org/records/13622420

Prior research has demonstrated a potential association between regional development and participation in the technology sector [12, 13]. Regions with greater access to education, strong economies, healthcare, infrastructure (including internet), and consumer technology tend to have a larger presence in the tech industry. This translates to a higher user base and increased activity levels on platforms like Stack Overflow, which holds true not only on a global scale but also within more granular levels, such as across states and cities [12]. An interesting dynamic exists in how users from different geographic locations interact within the community, where urban users tend to adopt a more assertive and solution-oriented approach, actively contributing information. In contrast, rural users primarily engage through inquiries and discussions, often incorporating personal experiences, expressions of gratitude, and conciliatory language [12]. Another separate study attempts to unveil more of this mosaic by going beyond simple quantitative enumeration [13], developing metrics to assess user participation, behaviour, and the value they contribute to the community, enabling operationalisations of constructs that are not readily available (e.g., user popularity, sentiment of posts, and readability), across various US states and cities [13]. Interestingly, users from rural states tend to post more frequently, curate content more actively, and produce content that is more readable, positive-toned, and with fewer errors. Those residing in cities without a strong tech presence show signs of lower participation, increased lurking behaviour, and a tendency to write longer code snippets. In fact, users from tech hubs were more inclined to present technical jargon and collaborative knowledge-sharing, often incorporating code examples, debugging tips, and relevant anecdotes. Conversely, rural users express a wider range of emotions, including hope, frustration, and satisfaction, alongside their manifold questions [13]. While previous studies have explored various aspects of user interaction on Stack Overflow, the impact of diversity and socio-technical factors on code snippet quality remains unexamined. This current research aims to bridge this knowledge gap.

## 2.2 Code Quality

The software development landscape is marked by rapid evolution, characterised by a diverse range of programming languages, frameworks, and paradigms [38, 39]. However, these progressions do not always adhere to practices that ensure software quality [40], thereby calling for quality control mechanisms. Software quality encompasses diverse characteristics like functionality, reliability, usability, efficiency, maintainability and portability, traditionally defined by standards like ISO/IEC 9126 and ISO/IEC 25010 [41, 42]. However, translating these abstractions into measurable metrics presents a challenge [43]. In response to this challenge, more pragmatic quality models have emerged, such as SQALE and Quamoco, bridging the abstraction-implementation gap by providing concrete frameworks for quality assessment [43]. Furthermore, the emergence of specialised maintainability models like SIG and Delta address particular concerns like technical debt [44, 45]. Readability models tackle the subjective nature of code comprehension and program understanding, as done by Scalabrino et al. [46] as well as Buse and Weimer [47]. Moreover, the concept of code quality may also be classified into either code smell, internal, or external quality attributes [48, 49]. There are also external quality attributes including compatibility, configurability, extensibility, and functionality [49]. In spite of these developments, their applicability to smaller, modular code snippets like those found on Stack Overflow remains a challenge. A multitude of studies have thus assessed the quality of code snippets hosted on the platform across various dimensions. Security has been a primary focus, with investigations into vulnerabilities in C/C++ and C# [50, 51, 52]. Code readability has also been studied, highlighting the influence of less nuanced factors like line length, whitespace, and formatting on both code quality and comprehension [53]. Studies have even explored compliance with style guides. For instance, Bafatakis et al. [9] examined whether Python snippets comply to Python Enhancement Proposal (PEP) 8 [54], whereas Campos et al. [10] investigated rule violations in JavaScript snippets and found that 82.9% of the violations pertain to stylistic issues. Ahmad and Ó Cinnéide [55] conducted an inquiry into the code cohesion of Java code snippets that were copied from Stack Overflow into GitHub projects. Results indicate that copied code snippets actually lowers cohesion, thereby making the codebase less robust [56]. Further research by Subramanian and Holmes on Java [8] illuminate the issue of code snippet incompleteness stemming from specific aspects (e.g., variable declarations) being frequently omitted.

While studies like Zerouali et al. [57] demonstrate the value of analysing multiple languages for specific quality dimensions, they have not yet been applied to Stack Overflow itself. Their work explored security vulnerabilities in JavaScript, Python, and Ruby packages within Docker Hub images, showcasing the feasibility of such multi-language analysis. On the other side of the coin, several works have holistically examined multiple dimensions under a unified umbrella. Meldrum et al. [11] investigated Java snippets across Stack Overflow answers through the lens of reliability and conformance to programming rules, readability, performance, as well as security. These findings reveal that the most occurring violation is related to readability, averaging approximately 10.5 readability violations per snippet, followed by 4.8 of reliability and conformance to programming rules, 0.5 of performance, and 0.01 of security. Geremia et al. [58] have done conducted a similar study, operationalising code quality to encompass snippets' total lines of code, cyclomatic complexity [59], and readability. Results underscore that these quality dimensions influence the likelihood of solutions being adopted by fellow developers. Pantiuchina et al. [60] delineate code quality across four dimensions: cohesion, coupling, readability, and complexity, while Gonzalez et al. [61] consider total refactors, code documentation, complexity and introduced issues (e.g., code smells, bugs, and vulnerabilities).

The aforementioned studies highlight the multifaceted nature of code quality, yet there remains a dearth of research that comprehensively assesses code quality on Stack Overflow across multiple dimensions and programming languages. Hence, our study endeavours to address this gap in the literature by offering a comprehensive snapshot of the platform's code quality landscape. Moreover, we investigate how code quality relates to census diversity indicators, enabling the identification of inherent regional factors that may drive (or deter) the quality of code snippets. We acknowledge the existence of other CQA platforms that allow code snippets in their posts, such as Cross Validated[5], Ask Ubuntu[6], or Code Review[7], yet our work focusses solely on Stack Overflow due to its wealth of information and peer-to-peer interactions that allow for subtle behavioural tendencies to be studied [62]. The following section delineates the methods that were employed in this study.

# 3 METHODS

In this section, we first reflect on established literature to define our study scope regarding code quality dimensions. Subsequently, we elucidate our strategy for data collection and processing, followed by outlining our methodology for selecting the languages to examine. We conclude this section by delineating our approach to address each research question.

## 3.1 Code Quality Definition

Considering the difficulty of defining *code quality* [63], literature has identified various dimensions that define such a construct. Specifically, we identified ten pertinent quality dimensions from prior studies (see Section 2), namely security, readability, performance, complexity, completeness, reusability, conformance to programming rules, reliability, coupling, and code cohesion. Several dimensions exhibit overlap upon closer examination. Meldrum et al. [11], examined "reliability" alongside "conformance to programming rules," but we observed that the former encompasses the latter since style guides prescribe consistent code presentation [9]. Boogerd and Moonen [64] identified 10 convention violations as significant predictors of faults in C code, implying a causal relationship between rule adherence and reliability. Due to this interrelationship, we merged "conformance to programming rules" and "reliability" under the latter. Similarly, we observed the phenomenon of "code smells," which describes potential design problems [65]. Recent works associate code smells with reliability due to their link to error-prone systems [66], as the intuition is that error-prone code is inherently less reliable. In light of this, we also consider our "reliability" dimension to include code smells. Another overlapping dimension is "performance" and "complexity," which exhibit an inverse relationship (i.e., as one increases, the other decreases) [67]. For example, larger codebases with more features have increased complexity, leading to performance degradation [67]. Similarly, finite state machines, widely used in game development, experience performance drops as complexity increases with more game behaviours [68].

[5] https://stats.stackexchange.com
[6] https://askubuntu.com
[7] https://codereview.stackexchange.com

As such, the decision was made to include "performance," subsequently representing "complexity" inside the former with the intuition that high-performing code snippets are less complex.

Certain dimensions are not applicable to individual code snippets and are designed for analysing large codebases. For instance, "code cohesion" measures the relatedness of elements within a class, while "coupling" assesses the strength of inter-file or inter-class dependencies [55, 60]. This inherent reliance on class-based object-oriented structure renders them inapplicable to individual code snippets, particularly in interpreted languages like Python and Ruby where classes and complex dependencies might be less prevalent. "Completeness" assesses whether snippets lack proper class declarations, which is unsuitable as SO posts are often missing crucial parts [8]. Finally, "reusability" requires a separate dataset to compare snippet reuse across platforms like F-Droid, GitHub, and BitBucket [69, 70]. Consequently, we focussed on four dimensions: reliability, readability, performance, and security, aligning with Meldrum et al. [11], being the most relevant study to our own pertaining to code quality on Stack Overflow across multiple quality dimensions. Notably, their study addresses the largest number of quality dimensions, thus ensuring the robustness of our subsequent analyses. As these dimensions were initially defined for Java, we sought to generalise them to other programming languages. Finally, these four dimensions closely align with the ones identified by Ndukwe et al. [71] following a set of semi-structured interviews of 50 practitioners: "Functionality", "Readability", "Efficiency", "Security" and "Reliability". Firstly, the "Functionality" dimension is encompassed within "Reliability", as the rationale is that reliable code inherently fulfils its intended function. Similarly, we consider "Efficiency" to be synonymous with "Performance," defining performant code as code that executes the desired task with minimal processing steps. Table 1 documents our chosen quality dimensions, the reasoning behind their selection, and their relevance to our case.

Table 1. Quality dimensions and their definitions

| Dimension | Definition | Motivation | Implications |
|---|---|---|---|
| Reliability | Code snippets ought not to be broken, introduce bugs, or prone to errors [11]. | It is assumed that users who contribute code snippets are responsible for providing accurate solutions, and thus their solutions should be free of errors in order to be deemed reliable [11]. | The reliability of code may enable us to discern whether certain users have a propensity for writing more dependable code than others. |
| Readability | Code snippets should adhere to the readability conventions specific to each programming language [9, 11]. | Adhering to established conventions ensure that code can be readily maintained in the future [11]. In contrast, if code is unreadable, it may introduce maintainability challenges or inhibit program comprehension [72]. | The readability of code may allow us to infer whether certain users tend to neglect programming conventions, thus rendering their code snippets difficult to comprehend. |
| Performance | Code snippets should be efficient and capable of performing the intended task with minimal processing steps [11]. | Software ecosystems that exhibit greater efficiency and less complexity would render them more scalable, meaning that they would perform well even when presented with larger amounts of input [73], as opposed to those that are less efficient. | Code snippet performance may enable us to determine whether certain users are more efficient in their implementation of solutions. |
| Security | Code snippets ought not to compromise security or present potential vulnerabilities that could be exploited [11, 50]. | In order to protect oneself from malicious attacks and hackers' intrusions, it is imperative to prioritise the avoidance of vulnerabilities and weaknesses when writing software systems [74]. | The security of code snippets may uncover certain users' inclinations to produce less secure code, jeopardising the security of others who incorporate these insecure snippets into their codebases. |

## 3.2 Data Collection and Processing

Our study leverages the Stack Overflow (SO) Database, sourced from the Stack Exchange Data Dump[8] superset seen in previous literature [22]. The relational database was stored on our local instance of

[8] https://archive.org/details/stackexchange

Microsoft SQL Server, containing a myriad of activities and raw measures across different tables, namely users' posts (questions and answers), comments, badges, and edits. Being the latest at the time of our investigation, we used the June 2022 release[9], which included 56,264,788 posts, 85,467,182 comments, and 22,796,157 edits from 17,922,426 unique users. Our replication package [27] presents the interactive database schema[10] and the total rows for each table[11]. To explore how diversity indicators influence code quality within the US, our study leverages census data from a prior study [12]. This data encompasses ten sets of census indicators for each state within the US, carefully refined from an initial pool of 465, thereby ensuring the scientific rigour and analytical merit of the chosen diversity indicators. These sets cover a multitude of diversity aspects, such as economy, infrastructure, and education. Table 2 below lists these census diversity indicators, and what diversity aspect each indicator represents.

Table 2. Selected state-level census indicators [12]

| **Diversity Aspect** | **Indicator Name** |
|---|---|
| Economy | *Per capita income in the past 12 months (in 2020 inflation-adjusted dollars)* |
| Education | *School Enrolment by Level of School for the Population 3 Years and Over: Enrolled up to Grade 12* |
| Ethnicity | *Race: White alone (% of all population)* |
| Gender | *Total Males*<br>*Total Females* |
| Health | *Females with health insurance coverage* |
| Poverty | *Gini index* |
| Science & Technology | *Households with one or more types of computing devices* |
| | *Internet Subscriptions in Household: With an Internet subscription* |
| Social Protection & Labour | *Total Unemployment* |

Several data processing measures were then undertaken. To focus our scope on code quality within the confines of the US, we needed to identify which users hail from this country. Therefore, the first objective was to determine users' country of origin, to which we excluded 14,174,843 users with null location fields, retaining data for 3,814,901 unique users. Second, inconsistencies in naming conventions were found, such as variations like "SF Bay Area" and "San Fran" both denoting San Francisco but treated as distinct locations. This inconsistency prevented automatic aggregation of user data with corresponding census diversity indicators at regional levels. Consequently, the inability to link individual regions with their specific diversity indicator hindered our investigation into how these indicators might influence code quality. To address such mismatches and enable aggregation, we explored 15 proprietary geocoding APIs from numerous GIS providers, namely ArcGIS, Bing Maps, GeocodeEarth, Geodict, Geograpy, Geonames, Geopy, Geotext, Google Maps, MapQuest, Nominatim, OpenCage, OpenStreetMaps, Photon, and TomTom. The selection considered factors such as suitability for the study, commercial availability, and insights from recent literature. Following a rigorous evaluation process across city, state, and country levels (see our replication package [27][12] for more details), MapQuest was ultimately selected to standardise all user "Location" entries and detect invalid entities. Several entities that were deemed as invalid include "Tatooine" (Star Wars) for Users.Id[13] 11896718 and "Valhalla" (Norse mythology) for Users.Id 10938505. Moreover, some users provided non-geographical locations like "Mostly at home" for Users.Id 2736499. As 67,318 invalid location entries were then discarded, we ensure that city, state, and country entries are entirely accurate. As a demonstration, the string "Manhattan, NYC" would return "New York City" as the city, "New York" as the state, and "United States" as the country. Our scope is limited to only users from the US, given the origin of Stack Overflow and the largest English-speaking country with English as the *lingua franca* of SE-related rigours [19]. It is also a leader in tech innovation [20, 21] and hosts the highest number of Stack Overflow users [12], making it a suitable choice for this study. Therefore, we lastly filtered out

[9] https://archive.org/details/stackexchange_20220606
[10] Replication package » Figures » SO Database Schema.html
[11] Replication package » Figures » SO Database Rows.png
[12] Replication package » Demonstrations » GeoAPI Evaluation Process
[13] Prefixes denote the table that a particular column is in. For instance, "Users.Id" refers to the column Id within the Users table of the Stack Overflow Database.

2,986,774 users whose locations point to other countries, yielding a final sample of 760,809 users with 8,951,562 total posts (6,596,105 from answers) that only encompasses US users.

To extract code snippets from their posts, we followed recent literature where we only examined code blocks in posts rather than in-line code snippets [11]. The rationale behind this is that in-line code snippets were typically added as part of text answers when users needed to provide an explanation. Code blocks, on the other hand, provided actual implementations to solve the problem at hand [11]. For instance, Figure 1 depicts the answer Posts.Id 72504547 about using C integer types in a Fortran program. Technical keywords such as *import* in the answer were inserted as in-line snippets (the red square), whereas example implementations were inserted as code blocks (blue square). Prior literature deemed it unfair to assess the quality of such in-line snippets [11], therefore our analysis looks only at code blocks.

You have to `import` the C datatypes from the module scope into each subroutine of your interface (remember that in Fortran, single subroutines have no external dependencies by default):

```
abstract interface
  function AlertMessage(atype, szFmt)
    import c_int32_t,c_ptr ! Import C datatypes from module into this interface
    integer(c_int32_t) :: atype
    type(c_ptr) :: szFmt
  end function
end interface
```

Figure 1. Example answer (Posts.Id 72504547) with in-line code snippet (red) and code block (blue)

Our analysis of the Stack Overflow database revealed that code snippets are stored within the Body column wrapped in HTML tags. In-line code snippets are enclosed by *<code>* tags, while full code blocks are surrounded by *<pre><code>* tags. We employed the string-matching algorithm of the Pandas[14] library to extract code blocks from answer posts, excluding in-line code snippets. We thus eliminated 2,884,589 answers that did not contain a code block, whilst keeping 3,711,516 answers having 6,213,554 distinct code blocks. As an example of the former, Figure 2 below depicts an answer with Posts.Id 72505278 remarking how transfer learning requires developers to exclude the last layer of a pretrained neural network. The given answer only provided a conceptualised theory without a code snippet, and hence was excluded from subsequent analysis. As one answer may contain several code blocks, we stored the extracted outcomes in a separate table called USCodes, where individual rows correspond to distinct code snippets represented by SnippetId. Each row also includes information about the parent answer, as well as authorship details such as OwnerUserId, city, and state (refer to our replication package for the database schema [27][15]).

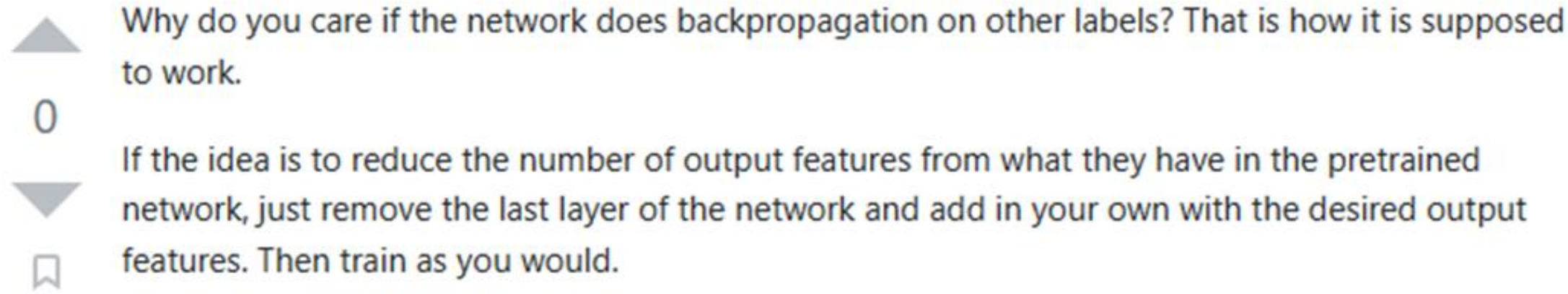

Figure 2. Example post that does not contain a code block (Posts.Id 72505278)

### 3.3 Language Inference and Selection

Following the extraction of code snippets from each answer, our subsequent task entails identifying the programming language(s) for further examination. This step was essential to facilitate pertinent pre-processing procedures such as insertion into classes [11] and exporting to appropriate file formats (e.g., .cpp for C++ or .js for JavaScript). Moreover, this knowledge enabled us to select specific languages representative of all users from the US. We initially considered harnessing the Posts.Tags column that often indicated the programming language in question (e.g., *'Java'*) and the technology being inquired (e.g., *'Maven'*) to determine which answers correspond to which questions. However, we encountered threads with multiple programming languages, rendering this approach unreliable. For example, Figure 3 depicts an answer with Posts.Id 38995602 outlining a workaround to display a dropdown menu. While the main question was tagged and posted in HTML, this particular answer was written in JavaScript.

[14] https://pandas.pydata.org/docs/reference/api/pandas.Series.str.contains.html
[15] Replication package » Figures » Snippets Database Schema.png

You can put a hidden field in the master page that holds the id of the current tab (parent `li`). Then set the value of that hidden field in the content pages.

Use the hidden field's value to determine which tab to highlight. This code assumes the parent `ul` has `id="MasterMenu"` and does **not** have `runat="server"`. Also that the hidden field has `id="CurrentTab"`.

```
function highlightTab() {
    var currentTab;

    //Get the id of the current tab from the hidden field
    currentTab = $('#CurrentTab').val();

    if (currentTab !== null) {
        //Remove the active-treeview class from all top level li elements
        $('#MasterMenu').children('li').removeClass('active-treeview');

        //Set the active-treeview class to the current tab
        $('#' + currentTab).addClass('active-treeview');
    }
}
```

Figure 3. Example answer (Posts.Id 38995602) in JavaScript

In light of this constraint, we pivoted towards Guesslang[16] to detect programming languages from plaintext. Guesslang has been proven to yield good effect in identifying code snippets on Stack Overflow [50, 51], and is the backbone for various developer tools such as Visual Studio Code[17] and Chameledit[18]. Prior to running the tool on our extracted code blocks, we did an empirical evaluation of 385 answers to ascertain its utility. Our evaluation revealed an accuracy of 84.67% (refer to our replication package [27] for full evaluation results[19]). Prior studies have established that accuracy exceeding 80% is considered acceptable [75]. Thus, our achievement of 84.67% ensured the reliability of our Guesslang application, enabling its application to our entire code snippet dataset. Results were then saved onto another column USCodes.GuesslangTag. While Table 3 below documents the identification of 54 distinct technologies after the tool has finished running, we also uncovered 4,551 snippets where Guesslang returned null values instead of the predicted technology. These were excluded, as they represented a small percentage (0.073%) of the entire pool of 6,213,554 snippets.

Table 3. 54 technologies identifiable by Guesslang

| Assembly | CSV | HTML | Markdown | R | Verilog |
|---|---|---|---|---|---|
| Batchfile | Dart | INI | Matlab | Ruby | Visual Basic |
| C | DM | Java | Objective-C | Rust | XML |
| C# | Dockerfile | JavaScript | OCaml | Scala | YAML |
| C++ | Elixir | JSON | Pascal | Shell | |
| Clojure | Erlang | Julia | Perl | SQL | |
| CMake | Fortran | Kotlin | PHP | Swift | |
| COBOL | Go | Lisp | PowerShell | TeX | |
| CoffeeScript | Groovy | Lua | Prolog | TOML | |
| CSS | Haskell | Makefile | Python | TypeScript | |

[16] https://github.com/yoeo/guesslang
[17] https://code.visualstudio.com/updates/v1_60#_automatic-language-detection
[18] https://gitlab.com/yoeo/chameledit
[19] Replication package » Assessments » Guesslang Sample Answers.xlsx

Within our new pool of 6,209,003 snippets, we discerned that several of these technologies are not programming languages, but rather data storage types, configuration files, as well as markup languages and UNIX-style commands such as sed (stream editor) or Shell scripts. For instance, JSON is a technology used to store data in nested key-value pairs, commonly returned as a GET or POST response from an API request. Another example pertains to INI, a file format mainly used for storing configuration settings in an application. Similar to JSON, the INI file format stores information in key-value pairs and thus has its own syntax so that the corresponding application can correctly read the written data.

Focussing primarily on programming languages, most prior research (see Section 2) omits analysis for configuration files, data storage formats, markup technologies, and UNIX-like commands. We posit that these technologies hold minimal relevance for code quality analysis, as applying our predetermined dimensions (see Table 1) to snippets that are not programming languages would be irrelevant. To first determine which of these predicted technologies belong to which category, we conducted a brief review of existing literature, because definitions of programming languages, data storage types, configuration files, markup languages, and UNIX-style commands can be subjective and thus vary from person to person. For instance, some might argue that HTML is a programming language, while others may see it as a markup language [76]. Similarly, some developers may not consider SQL to be a programming language since it cannot be used to create standalone applications. Yet the literature revealed that SQL is, in fact, a set-based, domain-specific, declarative programming language and thus should be categorised as such [77, 78]. Our replication package [27] documents our categorisation results, as well as their tallied occurrences[20].

We thus removed 891,764 configuration snippets, 87,149 data storage formats, 578,592 markup technologies, and 970,851 UNIX-like commands. This left us with a total of 3,680,647 snippets, encompassing 38 distinct programming languages. Despite having removed irrelevant snippets, the volume of data remains substantial, thereby bolstering the reliability of our findings. Previous research has demonstrated that even smaller code snippet pools can unearth novel insights [9, 11], and thus our study's downstream analyses based on this pool remain relevant. Next, we observed substantial variations in terms of their total number of snippets, hinting that languages may not be comparable between each other. For instance, SQL had 776,836 snippets, while Fortran only had 3,252. To compound this issue, certain languages were completely absent in particular states (e.g., no COBOL answers in Alabama). Another potential confounder is that we observed significant disparities in snippet counts across programming languages within individual states. For instance, Kansas had only 42 Scala snippets compared to 1,421 SQL and 951 JavaScript snippets. This phenomenon was prevalent among less popular languages like COBOL, Verilog, and DM, while widely used languages like SQL, JavaScript, and Python consistently exhibited higher snippet counts regardless of state.

To determine the languages for inclusion, we computed the mean number of snippets across all languages as a threshold, aligning with prior literature [79, 80]. Yielding an average of 96,859, 27 languages were excluded where their total snippets fell below this number, in turn retaining the top 11 languages. Sorted from most to least snippets, the top 11 languages are SQL (776,836 snippets), JavaScript (486,175), Python (329,396), Ruby (206,090), Java (184,903), C# (179,723), TypeScript (135,223), PHP (132,191), R (139,448), Groovy (121,668), and finally CoffeeScript (99,379). To conclude our funnelling process, our next objective was to ascertain which of these 11 languages were among the top 10 languages for each state. The rationale was that we would pick languages that offer comprehensive representation across all regions of the US. This was achieved by a simple quantitative analysis that tallied the number of code snippets in each language within each state. Our findings are documented in Table 4, where checkmarks denote whether the language is ranked within the top 10 languages in the corresponding state. Table 4 reveals that SQL, JavaScript, Python, Ruby, and Java are the only languages that are present in the top 10 languages of all 50 states, as well as the District of Columbia (refer to our replication package [27] for the tallied numbers[21]).

[20] Replication package » Results » Predicted Technologies.docx
[21] Replication package » Language Counts » Programming Languages - State.xlsx

Table 4. Prevalence of programming language for each state

| State | SQL | JavaScript | Python | Ruby | Java | C# | TypeScript | PHP | R | Groovy | CoffeeScript |
|---|---|---|---|---|---|---|---|---|---|---|---|
| Alabama | ✓ | ✓ | ✓ | ✓ | ✓ | ✓ |  | ✓ |  | ✓ |  |
| Alaska | ✓ | ✓ | ✓ | ✓ | ✓ | ✓ |  | ✓ | ✓ | ✓ | ✓ |
| Arizona | ✓ | ✓ | ✓ | ✓ | ✓ | ✓ | ✓ | ✓ |  | ✓ | ✓ |
| Arkansas | ✓ | ✓ | ✓ | ✓ | ✓ | ✓ | ✓ | ✓ |  | ✓ | ✓ |
| California | ✓ | ✓ | ✓ | ✓ | ✓ | ✓ | ✓ | ✓ | ✓ | ✓ | ✓ |
| Colorado | ✓ | ✓ | ✓ | ✓ | ✓ | ✓ | ✓ | ✓ | ✓ | ✓ |  |
| Connecticut | ✓ | ✓ | ✓ | ✓ | ✓ | ✓ |  | ✓ | ✓ |  | ✓ |
| Delaware | ✓ | ✓ | ✓ | ✓ | ✓ | ✓ | ✓ | ✓ |  | ✓ | ✓ |
| District of Columbia | ✓ | ✓ | ✓ | ✓ | ✓ | ✓ | ✓ | ✓ | ✓ | ✓ | ✓ |
| Florida | ✓ | ✓ | ✓ | ✓ | ✓ | ✓ | ✓ | ✓ |  | ✓ | ✓ |
| Georgia | ✓ | ✓ | ✓ | ✓ | ✓ | ✓ | ✓ | ✓ | ✓ | ✓ |  |
| Hawaii | ✓ | ✓ | ✓ | ✓ | ✓ | ✓ | ✓ | ✓ | ✓ | ✓ | ✓ |
| Idaho | ✓ | ✓ | ✓ | ✓ | ✓ | ✓ | ✓ | ✓ |  | ✓ | ✓ |
| Illinois | ✓ | ✓ | ✓ | ✓ | ✓ | ✓ | ✓ |  | ✓ | ✓ |  |
| Indiana | ✓ | ✓ | ✓ | ✓ | ✓ | ✓ | ✓ | ✓ |  | ✓ | ✓ |
| Iowa | ✓ | ✓ | ✓ | ✓ | ✓ | ✓ | ✓ | ✓ | ✓ | ✓ |  |
| Kansas | ✓ | ✓ | ✓ | ✓ | ✓ | ✓ | ✓ | ✓ |  | ✓ | ✓ |
| Kentucky | ✓ | ✓ | ✓ | ✓ | ✓ | ✓ | ✓ | ✓ | ✓ | ✓ | ✓ |
| Louisiana | ✓ | ✓ | ✓ | ✓ | ✓ | ✓ | ✓ | ✓ |  | ✓ |  |
| Maine | ✓ | ✓ | ✓ | ✓ | ✓ | ✓ | ✓ | ✓ |  | ✓ |  |
| Maryland | ✓ | ✓ | ✓ | ✓ | ✓ | ✓ | ✓ |  | ✓ | ✓ | ✓ |
| Massachusetts | ✓ | ✓ | ✓ | ✓ | ✓ | ✓ | ✓ | ✓ | ✓ | ✓ |  |
| Michigan | ✓ | ✓ | ✓ | ✓ | ✓ | ✓ | ✓ | ✓ | ✓ | ✓ |  |
| Minnesota | ✓ | ✓ | ✓ | ✓ | ✓ | ✓ | ✓ | ✓ | ✓ | ✓ | ✓ |
| Mississippi | ✓ | ✓ | ✓ | ✓ | ✓ | ✓ | ✓ | ✓ |  | ✓ | ✓ |
| Missouri | ✓ | ✓ | ✓ | ✓ | ✓ | ✓ | ✓ | ✓ | ✓ | ✓ |  |
| Montana | ✓ | ✓ | ✓ | ✓ | ✓ | ✓ | ✓ | ✓ | ✓ | ✓ | ✓ |
| Nebraska | ✓ | ✓ | ✓ | ✓ | ✓ | ✓ | ✓ | ✓ |  | ✓ | ✓ |
| Nevada | ✓ | ✓ | ✓ | ✓ | ✓ | ✓ | ✓ | ✓ |  | ✓ | ✓ |
| New Hampshire | ✓ | ✓ | ✓ | ✓ | ✓ | ✓ | ✓ | ✓ | ✓ | ✓ |  |
| New Jersey | ✓ | ✓ | ✓ | ✓ | ✓ | ✓ | ✓ | ✓ |  | ✓ | ✓ |
| New Mexico | ✓ | ✓ | ✓ | ✓ | ✓ | ✓ |  |  | ✓ | ✓ |  |
| New York | ✓ | ✓ | ✓ | ✓ | ✓ | ✓ | ✓ | ✓ | ✓ | ✓ | ✓ |
| North Carolina | ✓ | ✓ | ✓ | ✓ | ✓ | ✓ | ✓ | ✓ | ✓ | ✓ |  |
| North Dakota | ✓ | ✓ | ✓ | ✓ | ✓ | ✓ | ✓ | ✓ |  | ✓ |  |
| Ohio | ✓ | ✓ | ✓ | ✓ | ✓ | ✓ | ✓ | ✓ |  | ✓ |  |
| Oklahoma | ✓ | ✓ | ✓ | ✓ | ✓ | ✓ | ✓ | ✓ | ✓ |  |  |
| Oregon | ✓ | ✓ | ✓ | ✓ | ✓ | ✓ | ✓ | ✓ | ✓ | ✓ | ✓ |
| Pennsylvania | ✓ | ✓ | ✓ | ✓ | ✓ | ✓ | ✓ |  | ✓ | ✓ | ✓ |
| Rhode Island | ✓ | ✓ | ✓ | ✓ | ✓ |  | ✓ | ✓ |  |  |  |
| South Carolina | ✓ | ✓ | ✓ | ✓ | ✓ | ✓ | ✓ | ✓ |  | ✓ | ✓ |
| South Dakota | ✓ | ✓ | ✓ | ✓ | ✓ | ✓ | ✓ | ✓ |  | ✓ | ✓ |
| Tennessee | ✓ | ✓ | ✓ | ✓ | ✓ | ✓ | ✓ | ✓ | ✓ | ✓ | ✓ |
| Texas | ✓ | ✓ | ✓ | ✓ | ✓ | ✓ | ✓ | ✓ |  | ✓ | ✓ |
| Utah | ✓ | ✓ | ✓ | ✓ | ✓ |  | ✓ | ✓ | ✓ | ✓ | ✓ |
| Vermont | ✓ | ✓ | ✓ | ✓ | ✓ | ✓ | ✓ | ✓ | ✓ |  |  |
| Virginia | ✓ | ✓ | ✓ | ✓ | ✓ |  | ✓ | ✓ | ✓ | ✓ | ✓ |
| Washington | ✓ | ✓ | ✓ | ✓ | ✓ | ✓ | ✓ | ✓ | ✓ | ✓ | ✓ |
| West Virginia | ✓ | ✓ | ✓ | ✓ | ✓ | ✓ | ✓ | ✓ | ✓ | ✓ | ✓ |
| Wisconsin | ✓ | ✓ | ✓ | ✓ | ✓ | ✓ | ✓ |  | ✓ | ✓ |  |
| Wyoming | ✓ | ✓ | ✓ | ✓ | ✓ | ✓ | ✓ | ✓ | ✓ | ✓ |  |

In contrast, other languages do not demonstrate similar levels of ubiquity. For example, C# does not feature among the top 10 languages of Rhode Island, Utah, or Virginia. Similarly, TypeScript is absent from the top 10 languages of Alabama, Alaska, Connecticut, and New Mexico. Thus, given the pervasiveness of SQL, JavaScript, Python, Ruby, and Java, we decided to investigate their code snippets across the established quality dimensions. Selection of all five languages ensures a fair and balanced sample, guaranteeing comparable results across regions. However, these five programming languages might not reflect global popularity as revealed by the Stack Overflow Developer Survey conducted in the same year [81]. This limitation is discussed in Section 6. As a final measure to ensure the accuracy of Guesslang's language identification, we randomly sampled 385 code snippets, one from each question post tagged with a specific language (i.e., resulting in five sets of 385 snippets, one for each language). This sample was then manually verified to confirm the correctness of Guesslang's predictions, and the results are documented in Table 5 below. All five evaluated sets exhibited relatively high accuracy, around 87-94%, which ensures the validity of such predictions. Our replication package [27] contains the full results of the manual language checks[22].

Table 5. Guesslang evaluation results for each language

| Language | Question Tag Identifier | Accuracy |
|---|---|---|
| SQL | `<sql>` | 94.29% |
| JavaScript | `<javascript>` | 91.42% |
| Python | `<python>` | 89.61% |
| Ruby | `<ruby>` | 87.79% |
| Java | `<java>` | 90.91% |

### 3.4 Code Snippet Extraction

Following language selection, we seek to extract the code snippets. We applied filtering to our USCodes table, selecting rows where the GuesslangTag column matched the target language. For example, to filter SQL snippets, we isolated rows where the GuesslangTag equalled *"SQL"* before writing them as individual files. Each snippet's filename comprised a language prefix and its corresponding SnippetId, so that results could be easily linked back to their corresponding posts and users. For example, the filename *SQL_613.sql* refers to an SQL snippet having a SnippetId of 613.For dynamically-typed languages SQL, JavaScript, Python and Ruby [82], snippets were extracted without any additional pre-processing measures or snippet repairs (e.g., importing classes or deleting spaces), as doing so would confound our results [11]. Moreover, dynamically-typed languages afford greater flexibility in variable types and do not enforce strict type checking at compile time, therefore bypassing the need for enclosing code within public classes (such as in C or C++) [83]. When extracting Python snippets, Windows Defender identified 3 as Keylogger scripts [84], which are SnippetId 4214594 from Posts.Id 57747365, 6165888 from 65077377, and 6165943 from 65258604. We manually inspected these snippets to ensure safety, subsequently adding these snippets to our local machine's whitelist.

Java, on the other hand, is classified as a statically-typed language wherein variable types must be explicitly declared before usage [82, 83]. Following recent literature [11, 85], we firstly checked whether the code snippets contained *import*, *package* or *class* keywords. Should these keywords be present, we saved the file as it is. Otherwise, we encapsulated the snippets in a public class structure, with the class name containing the corresponding SnippetId. For instance, Figure 4**Error! Reference source not found.** illustrates SnippetId 561 enclosed in a public class structure before being saved as *Java_561.java*. Having prepared the data for analysis, Table 6 presents the methods employed to answer each RQ, which are then explored in detailed within the following subsections.

Table 6. Methods used to answer each RQ

| RQ# | Research Question | Method |
|---|---|---|
| RQ1 | What is the quality of code snippets across different programming languages on Stack Overflow? | Quantitative analysis was conducted via linting tools, followed by a detailed exploration how the identified violations for each language might shed light on users' habits on SE-related practices. |

[22] Replication package » Assessments » Guesslang Per Language

| RQ# | Research Question | Method |
|---|---|---|
| RQ2 | How do coding practices among US contributors differ across states and cities in the United States? | Violation densities for each region are calculated, defining total violations per logical line of code. A comparative analysis of these violation densities is then performed across different regions to identify variations in coding practices. |
| RQ3 | How do diversity indicators of US regions affect code quality? | Pearson's correlation analysis was employed to investigate the relationships between state-level diversity indicators and the violation densities of each state. At a 5% significance level, the null hypothesis was that there is no correlation, while the alternate hypothesis was that a correlation exists. |
| RQ4 | What themes are prevalent in code quality violations, and how do these themes differ across US regions? | Inductive content analysis was applied to code snippets from three representative states. This allowed us to delve into the latent characteristics associated with each state and code quality dimension. |

```
Java_561.java ×
D: > Java_561.java
1   public class Java_561 {
2       @RequestMapping(value = "/restaurant/add",method = RequestMethod.POST)
3       @ResponseBody
4       public String addRestaurantWebView(@RequestBody Restaurant restaurant){
5           System.out.println(restaurant.getRestaurantName());
6           this.restaurantService.addRestaurant(restaurant);
7          return null;
8       }
9
10  }
```

Figure 4. Example Java snippet (SnippetId 561) being enclosed in a public class structure

### 3.5 Code Snippet Quality (RQ1)

To assess code snippet quality, we initially considered using off-the-shelf automated static analysis tools such as SonarQube and Codacy, which offer quantifiable quality metrics and support for various programming languages [86]. However, we found that these tools are primarily designed for analysing software repositories with interdependent files and established continuous integration and continuous delivery (CI/CD) workflows [87]. These inherent design misalignments, coupled with the fact that these automated code review tools fundamentally employ linting tools as their underlying mechanisms [88], pivoted us to harness linting tools instead. For instance, Codacy was shown to employ ESLint and PMD as built-in code pattern checkers, all of which are linting tools[23]. Therefore, we employed numerous linting tools to operationalise code quality across the four dimensions of reliability, readability, performance, and security.

The majority of linting tools examined in our study are predominantly designed as command-line interface (CLI) utilities. Consequently, we redirected their outputs to text files in order to systematically store the results. Following this, we employed text mining techniques utilising regular expressions to extract both the SnippetIds and violation names. A number of tools exhibited the capability to detect a significant number of violations. To ensure clarity and focus we report only the ten most frequently detected violations identified by the tools for each dimension. Presenting an exhaustive list would be impractical, as some tools generate hundreds of violations for a single dimension. The definitions and examples for these violations are sourced from the respective tool documentation, unless otherwise specified. Additional results, such as definitions and examples[24], as well as tallied occurrences for each state and city[25], can be found in our replication package [27]. Our selection of linting tools was tailored to each research question and programming language, ensuring that the chosen tools could effectively identify issues related to the specified quality dimensions, and reflecting similar prior research. Some tools may be able to address multiple research questions (i.e., inspect more than one quality dimension), and be supplemented with plugins to further expand their scope. Our tools of choice are outlined in Table 7.

[23] https://docs.codacy.com/repositories-configure/configuring-code-patterns
[24] Replication package » Violations » Reliability to Security
[25] Replication package » Results » Tallied Occurrences » Reliability to Security

Table 7. Tools choice by dimension and language

| Programming Language | Tool Name | Dimension(s) | Violations to Check |
|---|---|---|---|
| SQL | SQLLint [25] | Reliability | 9 |
| | SQLFluff [89] | Readability | 59 |
| | SQLCheck [90] | Performance | 21 |
| | | Security | 7 |
| JavaScript | ESLint [10] | Reliability | 23 |
| | | Readability | 31 |
| | | Performance | 9 |
| | ESLint Security Plugin* [91] | Security | 14 |
| Python | flake8 [92] | Reliability | 33 |
| | | Readability | 52 |
| | | Performance | 8 |
| | flake8-bandit* [93] | Security | 62 |
| Ruby | Rubocop [94, 95] | Reliability | 90 – Lint category |
| | | Readability | 305 – Layout, Naming, and Style categories |
| | | Security | 5 – Security category |
| | Rubocop Performance* [94, 95] | Performance | 24 |
| Java | PMD [11, 94] | Reliability | 220 |
| | Checkstyle [11, 96] | Readability | 56 – Google checks |
| | SpotBugs [11] | Performance | 37 – Performance category |
| | | Security | 41 – Security and Malicious Code categories |

*) Plugins that work in conjunction with the original tool.

In total, there are 1,106 violations to check across all dimensions. Several implementation details are to be considered, given that each tool works differently. Firstly, flake8 and SQLFluff report violations using codes, requiring users to consult each tool's documentation for interpretation. Next, Meldrum et al. [11] used FindBugs for Performance and Security analysis, but due to its discontinuation, we opted for SpotBugs, its actively maintained successor. Like FindBugs, SpotBugs requires compiled code (.class) to identify violations. Moreover, Rubocop and SpotBugs automatically categorise violations, enabling us to quantify each dimension's violations by consulting the documentation. For instance, Rubocop's 'Lint' violation category is explicitly defined to scrutinise potential ambiguities and detect error-prone behaviour[26]. As this specific use-case aligns with our definition of reliability, this category would quantify the reliability dimension for Ruby. Each language's quality assessment is first conducted by analysing reliability, as our chosen tools in Table 7 can determine the proportion of parsable and unparsable snippets. Afterwards, the readability, performance and security dimensions are assessed. Determining the degree of unparsable snippets is crucial to account for potential inaccuracies in language prediction by Guesslang. Figure 5 exemplifies such a case, displaying SnippetId 104673.

*SQL_104673.sql*

```
RoomId = db.Database.ExecuteSqlCommand("SELECT RoomId FROM dbo.Rooms WHERE
RoomCategory = '" + roomCategory + "'");
return RoomId;
```

Figure 5. Example snippet SQL_104673.sql. Red text denotes SQL syntax

This snippet written in C# executes a database query using driver code, and contains embedded SQL syntax (highlighted in red). Consequently, Guesslang misidentified its language and returned SQL instead of C#. Similar limitations are further addressed in Section 6. Our study primarily reports and discusses the top 10 violations for each language, tool, and quality dimension because these represent over 75% of each language's violations within each dimension, except for the Ruby readability dimension, which covers only 55.99% (addressed further in Section 6). Refer to our replication package

[26] https://docs.rubocop.org/rubocop/cops.html#lint

[27] for the complete list of violations[27]. Aside from brevity, we posit that reporting the top 10 violations effectively presents the overall code quality scene for each region across all four dimensions.

Following tool execution, we rigorously examined all 1,106 violations for suitability in snippet-level analysis, employing the method conducted by Meldrum et al. [11]. This ensured each violation was indeed applicable for snippet-level analysis within Stack Overflow, as opposed to the file-level analysis typically used in traditional codebases. We subsequently found 21 violations that were only suitable for file-level code assessments, and thus not applicable to the snippet-level nature of Stack Overflow data. For instance, the violation *no-undef* by ESLint (JavaScript) flags undefined variables. However, variable definitions within the snippet itself may not be necessary if the question thread presents sufficient context. Figure 6 depicts such a case where the definitions of *array* and *value* are not assigned prior to being used, yet contributors can still grasp the main question being asked. Moreover, there are certain violations that were raised by our pre-processing steps when extracting and exporting snippets to their respective file extensions. For example the violation *ClassWithOnlyPrivateConstructorsShouldBeFinal* by PMD (Java) dictates that classes with private constructors should be declared as final, since such classes may not be subclassed either way. Consequently, private constructors encased in a public class structure—which is our pre-processing step as seen in **Error! Reference source not found.**—raises this v iolation. Details for these violations are documented in Table 8, and we excluded these violations from the subsequent analysis and findings [11] as they do not reflect a genuine issue within the actual snippets.

Figure 6. Example unassigned variable usage in JavaScript

Table 8. Omitted violations for each language and tool

| Language | Tool | Violation Name | Reason for Omission | Violating Snippets |
|---|---|---|---|---|
| JavaScript | ESLint | *no-undef* | Variable definitions may be omitted when the question thread provides sufficient context. | 280,894 |
| Python | flake8 | *F821 (undefined name)* | | 106,163 |
| Java | PMD | *UnusedPrivateField* | Irrelevant to analyse at snippet-level [11]. | 4,240 |
| | | *UnusedPrivateMethod* | | 4,301 |
| | | *UncommentedEmptyConstructor* | | 1,139 |
| | | *UncommentedEmptyMethodBody* | | 1,849 |
| | | *UnusedImports* | | 0 |
| | | *UseUtilityClass* | Violations caused by pre-processing steps [11]. | 25,233 |
| | | *ClassWithOnlyPrivateConstructorsShouldBeFinal* | | 2,285 |
| | | *CloneMethodMustImplementCloneable* | | 33 |
| | | *CloneMethodReturnTypeMustMatchClassName* | | 49 |
| | | *ProperLogger* | | 18 |
| | Checkstyle | *OuterTypeFilename* | Violations caused by pre-processing steps [11]. | 94,395 |
| | | *Indentation* | | 92,392 |
| | | *WhitespaceAround* | | 29,631 |
| | | *CommentsIndentation* | | 4,762 |
| | | *OneTopLevelClass* | Irrelevant to analyse at snippet-level [11]. | 3,373 |
| | | *JavadocMethod* | | 99 |
| | SpotBugs | *UPM_UNCALLED_PRIVATE_METHOD* | | 284 |

[27] Replication package » Results » Code Quality Violations (RQ1)

| | | |
|---|---|---|
| *URF_UNREAD_FIELD* | Irrelevant to analyse | 234 |
| *UUF_UNUSED_FIELD* | at snippet-level [11] | 126 |

### 3.6 Code Snippet Violation Densities (RQ2)

The fundamental essence of our work lies in examining regional variations, to reveal insights on how different states and cities vary in reliable, readable, performant, and secure coding practices. However, relying solely on the number of violations would be unfair as it might disproportionately penalise regions with lower activity, and therefore would yield fewer parsable snippets and fewer violations. Our preliminary investigation revealed that regions with greater technological engagement, like California and New York, yielded more parsable snippets and consequently, a higher number of violations across all dimensions and languages. This has the potential to skew insights in more rural states such as North Dakota or West Virginia. To avoid obscuring such findings, we adopted a "violation density" measure from prior literature, enabling us to uniformly compare the concentration of violations across regions with respect to all four dimensions and five languages [97, 98, 99]. Such a measure is usually computed as the ratio of the number of static analysis violations, divided by the product size (i.e., size of the code) [97]. Typically, lines-of-code (LOC) serves as the product size [98], leading to the intuitive interpretation that a density exceeding 1 indicates more than one violation per line on average. LOC takes into account all code elements, such as logical statements, comments, and blank lines. Yet this approach can be inaccurate for snippets containing numerous blank lines or comments, as counting these non-contributing lines as LOC would skew the density metric and eventually create an unfair assessment. Literature has thus suggested to use logical lines-of-code (LLOC) which only analyses actual lines of code, thereby excluding empty lines and comments [100]. However, **Error! Reference s ource not found.** readability violations often pertain to whitespace and comments, so therefore utilising LLOC for the readability dimension would render these violations invisible and obfuscate the nuances we were trying to uncover in the first place. In light of this dilemma, we will analyse violation densities with respect to LOC exclusively for the readability dimension, and use LLOC for the other three dimensions.

We harnessed a simple Python script to first tally the total LOC for each snippet, followed by a comprehensive parsing process with regular expressions to differentiate between single-line comments, multi-line comments, and whitespace within the code structure. Through this procedure, we effectively strip away extraneous elements, ultimately obtaining the desired LLOC of each snippet. Live demonstrations[28] and comment identifiers[29] for our regular expressions can be seen in our replication package [27]. Afterwards, we averaged the violation densities of individual code snippets authored by users from each city and state of the US. This approach provided us with a representative average violation density for each region, enabling us to conduct a comparative analysis to identify regional disparities in violation densities.

### 3.7 Code Snippet Relations (RQ3)

To investigate the relationships between various diversity factors and code quality across different dimensions and programming languages, we conducted Pearson's correlation analysis at the state level, employing a 5% significance level. This decision was guided by prior research [12] that indicated that only 179 of all US cities satisfy the necessary data quality assumptions. Conducting such an analysis at the city level would likely yield unreliable results due to an incomplete sample. Pearson's correlation analysis was employed due to its effectiveness in prior research [12, 34] and its robustness to normality deviations. Given our dataset encompasses 51 data points (50 states and the District of Columbia), it is suitable for our scope [101].

Following the calculation of average violation densities per state (used to answer RQ2), these values were merged with state-level diversity indicators obtained from a prior study [12] (Table 2). This process resulted in a comprehensive dataset, where each row represents a distinct state and each column houses state-level information: violation densities across different programming languages and code quality dimensions, alongside the corresponding census diversity indicators. Only correlation pairs with $p \leq$

[28] Replication package » Regex Demonstrations » Regex Demonstrations.txt
[29] Replication package » Regex Demonstrations » Comment Identifiers.png

0.05 were kept to ensure meaningful results. All correlation pairs and their corresponding *p*-values were computed using the Python package SciPy[30].

### 3.8 Code Snippet Themes (RQ4)

Quantitative exploration on all 50 US states and the District of Columbia was conducted for RQ2 and RQ3 to establish an understanding of code quality trends across the US. However, quantitative analyses alone may not provide sufficient depth to fully grasp the nuances of user-provided code snippets. Content analysis (CA) was thus selected as a confirmatory step, offering a qualitative perspective to complement prior quantitative findings. CA is a well-established method widely used in computer science and related fields (e.g., software engineering and information systems), serving as a valuable tool to unveil latent textual patterns that are otherwise often missed by simple quantitative measurements [12, 102]. Studies with similar design often harness CA using either inductive or deductive approaches [103]. Inductive analysis follows a bottom-up strategy, where researchers begin with the data and allow themes and categories to emerge organically, leading to theory development at the conclusion [104]. Conversely, deductive content analysis takes a top-down approach, where pre-existing theoretical frameworks guide the identification of codes and categories. Given the exploratory nature of our research, we opted for inductive CA. Our approach is centred specifically on code snippets within post bodies, excluding other content types like post paragraphs, user profiles, and temporal data. This focus on code snippets ensures consistency across the various code quality dimensions we examine, maintaining alignment with the findings from the three prior RQs. Furthermore, focussing on only code snippets mitigates biases that might rise from user attributes (e.g., age, interests, or profession), ensuring the impartiality of subsequent findings. Lastly, this approach draws inspiration from prior research that employed inductive CA on users' posts' paragraphs to uncover novel insights [13]. We thus apply the same technique to users' code snippets, offering a fresh perspective on user-generated content through direct analysis of the code instead.

Following established practices [105], this study defines three key units for content analysis: sampling unit, recording unit, and context unit. The sampling unit refers to each individual data instance that is chosen for examination. The recording unit is the text component that is subjected to categorisation during the coding phase, whereas the context unit denotes the text necessary to establish additional context for the recording unit [104]. Our study selected individual code snippets from answer posts as the sampling unit, complementing previous RQs and allowing thematic investigation embedded within user-provided code solutions. Additionally, answers are the central content of any CQA platform [106]. Naturally, the recording unit is the code snippets themselves, because they constitute the core of user solutions and are commonly found in both open-source and commercial software [11]. Finally, question threads are designated as the context unit. The rationale is that analysing code snippets within the context of a question thread may provide additional insights into how they contribute to (or deviate from) the central line of inquiry. A manual qualitative analysis of all code snippets across all states would be impractical, as all five languages tally to 1,983,400 snippets. Given our study's focus on regional variations in coding practices, we therefore employed a targeted sampling approach to select which states would be subjected to CA. This selection process considered factors relevant to both the technological landscape and workforce diversity within each state. Factors to consider are the tech workforce concentration, LOC and LLOC volume, the volume of tech job postings, tech job added, the economic impact of the tech sector, and Simpson's diversity index for that state – which served as a measure of racial and ethnic diversity among tech workers in a state [107]. Higher values indicate more even representation of otherwise underrepresented minorities (e.g., African American or Asian). Based on previous work [13] we also consider user base, urbanisation level, technology workforce concentration, R&D activity, and educational attainment. This ensures that our selection is not arbitrary, enabling a heterogeneous representation across manifold diversity facets. Subsequently, California, Utah, and North Dakota were selected as study sites due to their diverse profiles aligning with these key considerations. For instance, the state California boasted 1,487,864 tech jobs in 2022, whilst Utah had 123,450 and North Dakota only 13,272 within the same year (see Table 9 for details). This selection of states aligns with prior research [13], where these same geographical locations were chosen as study sites.

[30] https://scipy.org

Table 9. States to be subjected to content analysis

| State | Tech Employment | Avg. LOC per Snippet | Avg. LLOC per Snippet | Tech Business Concerns | Economic Impact to Tech Sector | Growth of Tech Job Openings | Simpson's Diversity Index* |
|---|---|---|---|---|---|---|---|
| California | 1,487,864 | 14.575 | 11.801 | 55,868 | 16.7% | +38,186 (+2.6%) | 0.649 |
| Utah | 123,450 | 15.379 | 12.532 | 9,286 | 10.0% | +5,130 (+4.3%) | 0.336 |
| North Dakota | 13,272 | 16.247 | 13.212 | 1,219 | 3.1% | +324 (+2.5%) | 0.216 |

* Calculated as $1 - D$ [108]; not to be confused with Simpson's Reciprocal Index $1/D$.

An interesting observation emerged when comparing LOC and LLOC within Table 9. California exhibited the lowest LOC and LLOC values, while North Dakota had the highest. This finding appears to align with established practices within the Stack Overflow community, which generally favours more concise and succinct code implementations [109]. Reflecting on the community's preference, California's code might be considered "better" despite having lower overall LOC and LLOC volumes. Following established practices [110], we adopted a probabilistic random sampling and selected 385 snippets for each state and each dimension. It is worth noting that our CA does not differentiate languages within each dimension, as findings in RQ1 revealed that violations largely correspond to the same thematic categories across all five programming languages. For instance, readability violations consistently involve aspects like code clutter and improper whitespace, while security vulnerabilities mainly revolve around unvalidated user inputs that could lead to injection attacks. This means that there would be 12 sets (4 dimensions × 3 states) where each contains 385 sample snippets to be subjected to CA. Specifically, we sampled only those where Guesslang's language identification predicted one of either SQL, JavaScript, Python, Ruby, or Java. Thus, for each dimension, there would be a total of 1,155 code snippets (385 for each state).

Within each dimension, the first two authors conducted a pilot reliability test to calibrate inter-coder consistency and subsequently draft the coding themes [111]. Development of the initial set of coding themes was supplemented by the use of Google's Gemini 1.5 Pro generative language model, chosen for its superior reasoning capabilities compared to similar models like ChatGPT [112]. It is important to emphasise that Gemini only served as a complementary aid, and that the final coding decisions remained the responsibility of the researchers. A 10% subsample was drawn for this purpose [111, 113], resulting in a total of 116 code snippets. The researchers then analysed the content of this subsample across two iterations of the pilot test [114], identifying emergent codes which would then form the first set of coding themes. The second iteration then refined this obtained set, merging similar themes, renaming themes for clarity, and making necessary adjustments to ensure a balance between generalisability and specificity. For instance, under the Security dimension, the themes *Exec Usage* and *Eval Usage* were merged into the theme of *Code Evaluation*, as both constitute parsing arbitrary strings as code. Similarly, the Performance dimension saw the introduction of *Usage of Inefficient Methods* to identify situations where faster methods can substitute certain implementations. Each emergent coding theme was assigned the value 1 if said theme was apparent within the code snippet, and 0 otherwise. A final coding theme *Not Coded* was added to facilitate code snippets that either could not be assigned to any of the emergent categories [22], or that did not raise any violations. Following each iteration, coders discussed differences to build consensus [113]. Lastly, Cohen's Kappa (κ) coefficient was used to calculate inter-coder reliability [115]. Table 10 documents how the coding themes evolved across both iterations as well as the resulting κ values. Our replication package [27] contains all reliability test results[31].

We employed pairwise chi-squared tests to ascertain thematic variations across user-generated content from the three states across all quality dimensions [13, 22]. The null hypothesis assumed uniformity in thematic prevalence across the states at $p < 0.05$. Finally, theme-specific frequency and percentage analyses quantitatively explicate the extent and direction of state-level divergences in code snippet violations.

[31] Replication package » Content Analysis (RQ4) » Reliability to Security

Table 10. Evolution of coding themes

| Dimension | 1st Iteration | | 2nd Iteration | | Final κ |
|---|---|---|---|---|---|
| | No. of Themes | Pilot κ | No. of Themes | Pilot κ | |
| Reliability | 19 | 0.762 | 17 | 0.823 | 0.839 |
| Readability | 14 | 0.772 | 13 | 0.796 | 0.825 |
| Performance | 7 | 0.784 | 8 | 0.845 | 0.859 |
| Security | 15 | 0.800 | 10 | 0.855 | 0.869 |

# 4 RESULTS

We present a holistic overview in Table 11 by showcasing the total number of parsable and unparsable code snippets for each language. Furthermore, for Java, we record the proportion of compilable snippets within its parsable pool.

Table 11. Code snippet breakdown

| Language | Initial Pool | Unparsable | Parsable | Uncompilable | Compilable |
|---|---|---|---|---|---|
| SQL | 776,836 | 418,850 | 357,986 | – | – |
| JavaScript | 486,175 | 145,750 | 340,425 | – | – |
| Python | 329,396 | 101,347 | 228,049 | – | – |
| Ruby | 206,090 | 91,646 | 114,444 | – | – |
| Java | 184,903 | 86,421 | 98,482 | 90,032 | 8,450 |

Despite our analysis being limited to 8,450 compilable Java snippets from the US, this sample size is likely sufficient to draw meaningful conclusions. Prior research by Meldrum et al. [11] achieved good results using 8,010 snippets which constitutes snippets from all countries (i.e., not only US). This suggests that our sample size that only encompasses those from the US can still yield robust findings. Nonetheless, we still acknowledge this small subset as a limitation (see Section 6). Table 12 provides a comprehensive overview of the total number of occurrences for each of the top 10 violations across all languages, tools, and dimensions. Certain entries may not reach 10, however, due to limitations of the specific tools. For example, SQLLint, used for evaluating SQL reliability, only identifies 9 violations, and thus only 9 entries are included in the table for that tool. For violations beyond the top 10, results are documented in our accompanying replication package [27][32].

## 4.1 Code Snippet Quality (RQ1)

This section presents results for RQ1: *What is the quality of code snippets across different programming languages on Stack Overflow?*

We first observe patterns in the reliability dimension. Given that code snippets on Stack Overflow are generally presumed to offer correct solutions and not intentionally misleading information [11], our definition for reliability is the code should not be broken, introduce bugs, or be prone to errors. Within SQL, only a small portion (10,090; 2.81%) were found to raise at least one reliability violation. From Table 12, the most common violation is *unmatched-parentheses*, which occurred 31.33% times, indicating that users tend to overlook whether parentheses are opened and closed properly when constructing queries. This was followed by *my-sql-invalid-create-option* (23.73%) and *hungarian-notation* (17.58%). The former hints that users tend to erroneously use the *CREATE* statement when constructing tables, whereas the latter indicates continued reliance on Hungarian notation despite its declining popularity in favour of standardised SQL conventions [116].

[32] Replication package » Code Quality Violations (RQ1)

Table 12. Violations' occurrences and definitions for each language, tool, and dimension.

| Lang. | Tool | Violation Name | Definition | Occurrences | % |
|---|---|---|---|---|---|
| *Reliability Dimension* | | | | | |
| SQL | SQLLint | *unmatched-parentheses* | Number of parentheses is unbalanced. | 3,612 | 31.33 |
| | | *my-sql-invalid-create-option* | A *CREATE* statement is being given an invalid option. | 2,735 | 23.73 |
| | | *hungarian-notation* | Queries contain the identifier *sp_* or *tbl_* | 2,027 | 17.58 |
| | | *missing-where* | A *WHERE* clause is absent from the *DELETE* statement. | 1,487 | 12.9 |
| | | *my-sql-invalid-alter-option* | An *ALTER* statement is being given an invalid option. | 972 | 8.43 |
| | | *my-sql-invalid-drop-option* | *DROP* statements are given an invalid option. | 383 | 3.32 |
| | | *invalid-limit-quantifier* | Non-numerical inputs are present inside *LIMIT* statements. | 228 | 1.98 |
| | | *my-sql-invalid-truncate-option* | *TRUNCATE* statements are given an invalid option. | 78 | 0.68 |
| | | *odd-code-point* | Queries contain unsupported code points. | 6 | 0.05 |
| JavaScript | ESLint | *no-prototype-builtins* | Usage of certain *Object.prototype* methods that can produce unexpected results. | 989 | 45.31 |
| | | *no-cond-assign* | Assignment operators occur in conditional expressions. | 522 | 23.91 |
| | | *no-global-assign* | Assignment to built-in global variables (e.g., *undefined*) | 215 | 9.85 |
| | | *no-unreachable* | Presence of statements located after the *return*, *throw*, *break*, and *continue* keyword. | 153 | 7.01 |
| | | *no-inner-declarations* | Declaration of functions and variables in nested blocks. | 94 | 4.31 |
| | | *no-dupe-keys* | Duplicate keys within object literals. | 45 | 2.06 |
| | | *valid-typeof* | Typos in string literals within *typeof* function (e.g., "strnig" instead of "string") | 33 | 1.51 |
| | | *no-control-regex* | Use of escape characters (e.g., \x00 or \x1F) in regular expressions. | 35 | 1.60 |
| | | *no-const-assign* | Reassigning a *const* variable. | 25 | 1.15 |
| | | *no-async-promise-executor* | Use of *async* function with promise executors. | 26 | 1.19 |
| Python | flake8 | *F405 (name may be undefined)* | Seemingly-undefined variable usage ambiguously imported with the wildcard *import* statement. | 20,182 | 60.61 |
| | | *W605 (invalid escape)* | Usage of a backslash-character pair that does not form a valid escape sequence. | 6,323 | 18.99 |
| | | *E722 (bare except)* | Usage of a bare except clause. | 2,230 | 6.70 |
| | | *F706 (return outside function)* | Usage of *return* statement outside functions. | 2,111 | 6.34 |
| | | *F722 (forward annotation error)* | Syntax error in type annotation (type hinting). | 1,098 | 3.30 |
| | | *F704 (yield outside function)* | Usage of *yield* statement outside functions. | 287 | 0.86 |
| | | *F632 (is used for literals)* | Comparison of string, byte, or integers using *is* or *is not* keywords. | 225 | 0.68 |
| | | *F823 (undefined local)* | References of local variables prior to their definition. | 125 | 0.38 |
| | | *F633 (print redirection)* | Usage of >> operator is used with the *print()* function. | 184 | 0.55 |
| | | *F701 (break outside loop)* | Usage of *break* keyword outside loops. | 114 | 0.34 |
| Ruby | Rubocop | *UselessAssignment* | Unused variables within local scopes. | 19,352 | 49.92 |
| | | *Void* | Usage of operators, variables, literals, lambdas, and nonmutating methods in void context. | 5,955 | 15.36 |
| | | *NumberConversion* | Number conversions with the *to_i*, *to_f*, *to_r*, or *to_c* methods. | 2,724 | 7.03 |
| | | *UnusedBlockArgument* | Block arguments are not utilised anywhere in the block. | 1,981 | 5.11 |
| | | *UnusedMethodArgument* | Method arguments are not utilised within that method. | 1,214 | 3.13 |
| | | *MultipleComparison* | Comparisons of three or more values within a single line. | 558 | 1.44 |

| Lang. | Tool | Violation Name | Definition | Occurrences | % |
|---|---|---|---|---|---|
| | | *AmbiguousBlockAssociation* | A parameter is passed in a method without parentheses. | 504 | 1.30 |
| | | *AssignmentInCondition* | If/while/until blocks contain an assignment operator (=) rather than the equality operator (==). | 462 | 1.19 |
| | | *TopLevelReturnWithArgument* | A non-empty return statement is used in the top-level object. | 429 | 1.11 |
| | | *AmbiguousOperator* | Ambiguous operators are found in the first argument of a method invocation without parentheses. | 390 | 1.01 |
| Java | PMD | *LocalVariableCouldBeFinal* | A local variable assigned only once is not declared as *final*. | 169,031 | 31.88 |
| | | *MethodArgumentCouldBeFinal* | A method argument is not declared as *final*. | 139,483 | 26.31 |
| | | *SystemPrintln* | Usage of *System.out.print* or *System.err.print* instead of logging statements. | 52,108 | 9.83 |
| | | *LawOfDemeter* | A class is interacting with the internals of related objects or their collaborators (tight coupling). | 24,025 | 4.53 |
| | | *ImmutableField* | Private non-final fields whose value never changes are not marked as *final*. | 17,839 | 3.36 |
| | | *ConstructorCallsOverridableMethod* | A non-private constructor calls an overridable method from within a class. | 11,048 | 2.08 |
| | | *AvoidInstantiatingObjectsInLoops* | New objects are instantiated within loops. | 6,455 | 1.22 |
| | | *DoNotUseThreads* | Threads are directly used within J2EE components (Java 2 Platform, Enterprise Edition) | 6,179 | 1.17 |
| | | *UnusedLocalVariable* | A local variable is declared and/or assigned, but not used. | 6,017 | 1.14 |
| | | *TestClassWithoutTestCases* | Test classes do not contain any test cases nor test methods. | 5,569 | 1.05 |
| *Readability Dimension* | | | | | |
| SQL | SQLFluff | *LT02 (layout.indent)* | Incorrect indentation in the query. | 176,834 | 15.27 |
| | | *LT01 (layout.spacing)* | Inappropriate spacing, such as excessive or trailing whitespace. | 150,405 | 12.99 |
| | | *LT09 (layout.select_targets)* | Multiple *SELECT* targets are placed on the same line, except when there's only one target. | 130,087 | 11.23 |
| | | *CP02 (capitalisation.identifiers)* | Capitalisation inconsistencies of unquoted identifiers, such as column names. | 105,311 | 9.09 |
| | | *AL01 (aliasing.table)* | Tables are aliased without using the *AS* keyword. | 99,896 | 8.63 |
| | | *CP01 (capitalisation.keywords)* | Capitalisation inconsistencies of keywords, such as *SELECT* or *FROM*. | 57,373 | 4.95 |
| | | *AM04 (ambiguous.column_count)* | A query generates an unspecified number of result columns. | 55,382 | 4.78 |
| | | *LT05 (layout.long_lines)* | Excessively long queries without a line break, leading to complex and hard-to-read queries. | 45,886 | 3.96 |
| | | *AL03 (aliasing.expression)* | Column expressions are implicitly aliased without using the *AS* keyword. | 32,573 | 2.81 |
| | | *RF03 (references.consistent)* | Inconsistent references as prefixes. | 28,894 | 2.50 |
| JavaScript | ESLint | *no-unused-vars* | Unused variable, contributing to code bloat. | 94,604 | 90.12 |
| | | *no-extra-semi* | Semicolons that are unnecessarily written. | 2,857 | 2.72 |
| | | *no-redeclare* | Redeclarations uses *var* keyword. | 2,559 | 2.44 |
| | | *no-useless-escape* | Escape character usage (backslash) when there is nothing to escape. | 2,514 | 2.40 |
| | | *no-unused-labels* | Declared but unused labels. | 845 | 0.81 |
| | | *no-constant-condition* | Constant expressions in conditional statements. | 379 | 0.36 |
| | | *no-irregular-whitespace* | Unicode whitespace characters are used yet not visually distinguishable from regular spaces. | 353 | 0.34 |
| | | *no-empty* | Occurrences of empty block statements. | 262 | 0.25 |
| | | *no-unexpected-multiline* | Usage of certain operators (such as *&&* and *:*) consecutively within a single line. | 175 | 0.17 |
| | | *no-shadow-restricted-names* | JavaScript global objects and restricted identifiers are renamed. | 141 | 0.13 |

| Lang. | Tool | Violation Name | Definition | Occurrences | % |
|---|---|---|---|---|---|
| Python | flake-8 | *E231 (no space around delimiters)* | Absence of whitespace after the comma, colons, or semicolons. | 267,156 | 29.46 |
| | | *E501 (line too long)* | Lines that span longer than 79 characters. | 101,985 | 11.25 |
| | | *E111 (indentation invalid multiple)* | Indentation used is not a multiple of four spaces. | 74,068 | 8.17 |
| | | *E302 (no blank line after functions)* | Two blank lines are not found between functions and classes declared in the global scope. | 65,804 | 7.26 |
| | | *W291 (trailing whitespace)* | Presence of trailing whitespace after the final character in a line. | 61,904 | 6.83 |
| | | *E225 (no space around operator)* | Absence of a single whitespace around comparison operators. | 53,920 | 5.95 |
| | | *E251 (space around default keywords)* | Presence of spaces before or after the "=" sign in a function definition. | 47,977 | 5.29 |
| | | *E261 (two spaces before inline comment)* | Presence of insufficient spacing before inline comments. | 33,228 | 3.66 |
| | | *E265 (no space after pound)* | Absence of whitespace before the pound sign and the comment itself within block comments. | 28,457 | 3.14 |
| | | *E128 (continuation under-indented)* | Insufficient indentation for continuation lines in a supposedly-single line statement. | 16,756 | 1.85 |
| Ruby | Rubocop | *Style/StringLiterals* | Double quotation marks (") are used to construct a string literal. | 74,910 | 13.56 |
| | | *Style/MethodCallWithArgsParentheses* | Absence of parentheses in method calls containing parameters. | 45,401 | 8.22 |
| | | *Style/DocumentationMethod* | Absence of documentation comments for public methods. | 34,826 | 6.30 |
| | | *Style/HashSyntax* | Inconsistent use of the Ruby 1.9 syntax (e.g., {a: 1}) when the hash's keys are of *Symbol* type. | 33,779 | 6.11 |
| | | *Layout/SpaceAroundOperators* | Absence of whitespaces around arithmetic, assignment, comparison, and logical operators. | 23,010 | 4.16 |
| | | *Layout/TrailingWhitespace* | Presence of trailing whitespace. | 22,406 | 4.05 |
| | | *Style/Documentation* | Absence of top-level documentation of classes and modules, except for empty classes and modules. | 21,509 | 3.89 |
| | | *Style/TopLevelMethodDefinition* | Top-level method declarations are not organised within their appropriate classes. | 19,622 | 3.55 |
| | | *Layout/IndentationWidth* | Indentation does not comply to a multiple of 2 spaces. | 17,131 | 3.10 |
| | | *Layout/SpaceAfterComma* | Comma not followed by any space. | 16,854 | 3.05 |
| Java | Checkstyle | *MissingJavadocMethod* | Absence of Javadoc comments for a method or constructor. | 55,304 | 17.45 |
| | | *LeftCurly* | Placement of left curly braces ("{") for code blocks is not at the end of line. | 46,318 | 14.61 |
| | | *LineLength* | Lines that span longer than 100 characters. | 27,312 | 8.62 |
| | | *ParenPad* | Inconsistent padding around parentheses in method or constructor declarations, method calls, and conditionals. | 26,556 | 8.38 |
| | | *EmptyLineSeparator* | Absence of empty line separators before package, imports, fields, methods, classes, and static/instance initialisers. | 20,869 | 6.59 |
| | | *CustomImportOrder* | Import statements in a source file do not follow any specified order. | 19,011 | 6.00 |
| | | *NeedBraces* | Absence of braces inside single-line statements. | 16,581 | 5.23 |
| | | *MemberName* | Instance variable names do not match to the regular expression *^[a-z][a-zA-Z0-9]*$*. | 14,525 | 4.58 |
| | | *AbbreviationAsWordInName* | Variable, method, or class names contain abbreviations that are treated as separate words. | 12,037 | 3.80 |
| | | *AvoidStarImport* | *import* statements uses the * notation. | 10,081 | 3.18 |

| Lang. | Tool | Violation Name | Definition | Occurrences | % |
|---|---|---|---|---|---|
| *Performance Dimension* | | | | | |
| SQL | SQLCheck | *SELECT ** | Query does not return an explicit number of columns. | 48,257 | 42.85 |
| | | *Spaghetti Query Alert* | Presence of large spaghetti queries. | 22,592 | 20.06 |
| | | *UNION Usage* | Usage of *UNION* instead of *UNION ALL.* | 19,063 | 16.93 |
| | | *Implicit Column Usage* | Wildcard usage and unnamed columns in querying. | 8,564 | 7.60 |
| | | *Generic Primary Key* | A generic name of primary key column (e.g., Id) for tables exist. | 5,124 | 4.55 |
| | | *Metadata Tribbles* | Mixing of metadata with data. | 4,151 | 3.69 |
| | | *Index Attribute Order* | Query attributes are not in the same order as the index attributes. | 1,870 | 1.66 |
| | | *Multi-Valued Attribute* | An attribute has multiple values for a single instance of an entity. | 1,214 | 1.08 |
| | | *ORDER BY RAND Usage* | Query selects a randomised set of rows from the result. | 760 | 0.68 |
| | | *DISTINCT & JOIN Usage* | Query contains the *DISTINCT* and *JOIN* clauses, instead of *EXISTS*. | 563 | 0.50 |
| JavaScript | ESLint | *no-redeclare* | Redeclaration of variables using the *var* keyword. | 2,559 | 93.57 |
| | | *no-debugger* | Usage of debugger statements, which stops execution at the current point in the code. | 94 | 3.44 |
| | | *no-fallthrough* | Absence of *break* keyword within switch-case statements. | 51 | 1.87 |
| | | *no-misleading-character-class* | Usage of certain character classes (such as Emojis) in regular expressions. | 14 | 0.51 |
| | | *for-direction* | Presence of infinite for-loops due to a wrong loop direction. | 9 | 0.33 |
| | | *require-yield* | Usage of *yield* keyword outside functions. | 5 | 0.18 |
| | | *no-useless-backreference* | Unnecessary backreferences in regular expressions. | 2 | 0.07 |
| | | *no-invalid-regexp* | Usage of invalid regular expression patterns (e.g., "[" or ".") in *RegExp* constructors. | 1 | 0.04 |
| Python | flake8 | *F401 (unused imports)* | Presence of imported but unused modules. | 18,933 | 51.73 |
| | | *F841 (unused variable)* | Defined but unused local variables. | 6,555 | 17.91 |
| | | *E402 (import placement)* | Module-level imports are not located at the top of the file. | 4,923 | 13.45 |
| | | *F403 (wildcard imports)* | Usage of the *from module import ** statement (known as wildcard import). | 4,095 | 11.19 |
| | | *E712 (comparison to bool)* | Comparison of a variable to the Boolean value *True*. | 1,317 | 3.60 |
| | | *F811 (redundant imports)* | Modules are imported multiple times in the same namespace. | 774 | 2.12 |
| | | *F842 (unused annotation)* | Defined but unused local, annotated variables. | 3 | 0.01 |
| | | *F622 (multiple starred expressions)* | Use of multiple starred expressions in assignment statements, such as *a, *b = *c, d*. | 1 | 0.00 |
| Ruby | Rubocop Performance | *RegexpMatch* | Usage of conventional *match* methods within regular expressions. | 466 | 28.68 |
| | | *StringReplacement* | Usage of *gsub()* method with string literals, which should be replaced by *tr* or *delete*. | 238 | 14.65 |
| | | *RedundantBlockCall* | Usage of *&block* and *block.call* in generators, which should be replaced by *yield*. | 168 | 10.34 |
| | | *RedundantMatch* | Usage of *Regexp#match* or *String#match* instead of using =~. | 131 | 8.06 |
| | | *TimesMap* | Usage of *times.map*, where an explicit array creation is a recommended substitute. | 124 | 7.63 |
| | | *CompareWithBlock* | Usage of *sort { \|a, b\| a.foo < ⇒ b.foo }* which should be replaced by *sort_by(&:foo).* | 64 | 3.94 |
| | | *RedundantMerge* | Usage of *Hash#merge!* – which should be replaced by *Hash#[]=.* | 62 | 3.82 |
| | | *RangeInclude* | Usage of *Range#include?* and *Range#member?,* which should be replaced with *Range#cover?* | 59 | 3.63 |
| | | *BindCall* | Usage of *bind(obj).call(args, ...)* instead of *bind_call(obj, args, ...).* | 51 | 3.14 |

| Lang. | Tool | Violation Name | Definition | Occurrences | % |
|---|---|---|---|---|---|
| | | *Casecmp* | Case-insensitive string lexicographical comparisons are not done with *casecmp*. | 49 | 3.02 |
| Java | SpotBugs | *SBSC_USE_STRINGBUFFER_CONCATENATION* | Presence of string concatenation using the '+' operator in a loop. | 58 | 25.89 |
| | | *SIC_INNER_SHOULD_BE_STATIC_ANON* | An anonymous inner class is not declared as static. | 46 | 20.54 |
| | | *SS_SHOULD_BE_STATIC* | A class contains an instance final field that is initialised to a compile-time static value. | 25 | 11.16 |
| | | *SIC_INNER_SHOULD_BE_STATIC* | An inner class (not necessarily anonymous) is not declared as static. | 21 | 9.38 |
| | | *DM_NEXTINT_VIA_NEXTDOUBLE* | The *nextInt()* method is not used in random number generation. | 19 | 8.48 |
| | | *DM_BOXED_PRIMITIVE_FOR_PARSING* | A boxed primitive is created from a *String*, just to extract the unboxed primitive value. | 18 | 8.04 |
| | | *DM_STRING_CTOR* | The constructor of the *java.lang.String* class is used to create a new string object. | 18 | 8.04 |
| | | *DM_GC* | Explicit calls to garbage collection. | 4 | 1.79 |
| | | *DM_STRING_VOID_CTOR* | Creation of an empty strings using new *String()* constructor. | 3 | 1.34 |
| | | *DM_STRING_TOSTRING* | Usage of *String.toString()* operation. | 3 | 1.34 |
| *Security Dimension* | | | | | |
| SQL | SQLCheck | *SELECT ** | Query does not return an explicit number of columns. | 48,257 | 38.22 |
| | | *NULL Usage* | *NULL* is used to conduct comparison operations. | 47,652 | 37.74 |
| | | *Spaghetti Query Alert* | Presence of large spaghetti queries. | 22,592 | 17.89 |
| | | *String Concatenation* | Usage dynamic of string concatenation on querying. | 3,695 | 2.93 |
| | | *Imprecise Data Type* | Usage of *FLOAT* instead of *DOUBLE* and *REAL* when inserting values for the database. | 3,565 | 2.82 |
| | | *Readable Passwords* | Password are stored in plain-text, or is decipherable by a cryptographic function. | 499 | 0.40 |
| JavaScript | ESLint Security Plugin | *object-injection* | Presence of square bracket notation in objects as a left- or right-hand assignment operand. | 80,582 | 94.08 |
| | | *non-literal-regexp* | Dynamic variable usage in *RegExp()* functions, rendering expressions non-literal. | 1,363 | 1.59 |
| | | *non-literal-fs-filename* | Usage of *module*[33] that is made with a dynamic filename argument. | 916 | 1.07 |
| | | *unsafe-regex* | Usage of unsafe regular expressions, which may block the event loop. | 800 | 0.93 |
| | | *eval-with-expression* | Usage of *eval(),* which executes a string of characters as code instead. | 798 | 0.93 |
| | | *non-literal-require* | Non-literals (e.g., variables or user input) being passed into the *require()* mechanism. | 793 | 0.93 |
| | | *new-buffer* | A new *Buffer* object is created with a non-literal value for its length. | 241 | 0.28 |
| | | *possible-timing-attacks* | Usage of insecure comparisons (==, !=, !== and ===) to check input sequentially. | 99 | 0.12 |
| | | *child-process* | A call to the *child_process* module is made with a non-literal argument. | 56 | 0.07 |
| | | *bidi-characters* | Usage of bidirectional characters, such as *RLO*[34] and *PDI*[35]. | 2 | 0.00 |
| Python | flake8-bandit | *S311 (pseudorandom)* | Usage of standard pseudo-random generators, specifically those from the *random* standard library. | 6,708 | 23.40 |
| | | *S101 (assert usage)* | Usage of *assert* keyword as a safeguard mechanism. | 4,194 | 14.63 |
| | | *S113 (request without timeout)* | A request with the *requests* library is made without specifying a *timeout*. | 3,143 | 10.96 |

[33] https://nodejs.org/api/fs.html
[34] https://www.compart.com/en/unicode/U+202E
[35] https://www.compart.com/en/unicode/U+2069

| Lang. | Tool | Violation Name | Definition | Occurrences | % |
|---|---|---|---|---|---|
| | | *S603 (subprocess without shell)* | Usage of subprocess module without explicitly setting *shell=True* for command execution. | 1,785 | 6.23 |
| | | *S607 (process spawned with partial path)* | Filesystem paths do not start at the root (i.e., lack of a leading forward slash character). | 1,470 | 5.13 |
| | | *S404 (subprocess imports)* | Importation of the *subprocess* standard library to spawn child processes. | 1,449 | 5.05 |
| | | *S310 (urlopen with file scheme)* | Usage of the *urlopen()* function from the *urllib* standard library. | 1,427 | 4.98 |
| | | *S105 (hardcoded passwords)* | String literals resemble hardcoded passwords. | 739 | 2.58 |
| | | *S108 (hardcoded temp)* | Usage of hardcoded temporary file or directory paths. | 685 | 2.39 |
| | | *S605 (process spawned with shell)* | Calls that start a process with a shell in an unsafe manner. | 589 | 2.05 |
| Ruby | Rubocop | *Open* | Usage of *Kernel#open* and *URI.open* with dynamic data (i.e., unsanitised input). | 561 | 53.94 |
| | | *Eval* | Usage of *Kernel#eval* and *Binding#eval* methods, which evaluates strings as code. | 272 | 26.15 |
| | | *YAMLLoad* | Usage of YAML class methods to load YAML objects. | 161 | 15.48 |
| | | *MarshalLoad* | Usage of *Marshal* class methods, especially the methods *load* and *restore*. | 24 | 2.31 |
| | | *JSONLoad* | Usage of *JSON* class methods, especially the methods *load* and *restore*. | 22 | 2.12 |
| Java | SpotBugs | *MS_SHOULD_BE_FINAL* | Public static fields are not declared as final. | 65 | 40.12 |
| | | *MC_OVERRIDABLE_METHOD_CALL_IN_CONSTRUCTOR* | An overridable (non-static, non-final, public) method is called within a constructor. | 38 | 23.46 |
| | | *MS_FINAL_PKGPROTECT* | A mutable public static field is neither declared as protected nor final. | 12 | 7.41 |
| | | *MS_EXPOSE_REP* | A public static method returns a reference to an array that is part of the static state of the class. | 10 | 6.17 |
| | | *EI_EXPOSE_REP* | A mutable internal object is exposed by pointing a reference to it – from an object or a publicly-accessible field. | 10 | 6.17 |
| | | *EI_EXPOSE_REP2* | Code stores a reference to an externally mutable object into the internal representation of the object. | 10 | 6.17 |
| | | *MS_PKGPROTECT* | Presence of a mutable, non-protected static field. | 9 | 5.56 |
| | | *DP_DO_INSIDE_DO_PRIVILEGED* | Method invocations require a security permission check. The invocation needs to occur inside a *doPrivileged* block. | 3 | 1.85 |
| | | *REFLC_REFLECTION_MAY_INCREASE_ACCESSIBILITY_OF_CLASS* | Usage of reflection to increase accessibility of classes, methods or fields. | 2 | 1.24 |
| | | *DMI_CONSTANT_DB_PASSWORD* | A database password is hardcoded into the code as a string. | 2 | 1.24 |

JavaScript snippets showed much fewer reliability violations compared to SQL where only 1,768 (0.52%) managed to flag at least one. The most common problem was *no-prototype-builtins*, accounting for 45.30% of all violations, possibly due to habits formed during the migration to ECMAScript 5.1[36]. This was followed by *no-cond-assign* (23.91%) and *no-global-assign* (9.85%). The former signals a proclivity for assignments within conditional blocks. The latter, while less frequent, is somewhat more severe: inadvertently modifying built-in global variables (e.g., *undefined*) which may have detrimental consequences and lead to unanticipated behaviour.

Python showed a moderate amount of reliability violations with 10,234 (4.49%) across all 228,049 parsable snippets. *F405 (name may be undefined)* occurred the most, accounting for 60.61% of all violations, suggesting the usage of variables prior to its definition – likely due to the usage of wildcard, implicit imports. *W605 (invalid escape)* followed closely at 18.99% as well as *E722 (bare except)* at 6.7%. These figures hint at the widespread misuse of escape characters and omission of proper exception handling in try-except blocks. The latter may obscure the true cause of errors and thus make the code behave unexpectedly [11].

Ruby code exhibited more reliability violations at 24,302 (21.23%) across all 114,444 snippets. *UselessAssignment* occurred the most (49.92%) followed by *Void* (15.36%). The former suggests a widespread adoption of declaring variables that ultimately remain unused, whereas the latter is more severe; such practices would mask underlying program logic due to syntax errors within the enclosing code block. *NumberConversion* accounted for 7.03% of all violations, suggesting users tend to use less resilient number conversion techniques, leading to unexpected numerical outcomes[37].

Finally, 84.36% of Java snippets (83,080 of 98,482 parsable) raised at least one violation. *LocalVariableCouldBeFinal* is the most prevalent – covering 31.88% of all violations – hinting a widespread practice to avoid declaring variables as *final* despite being assigned only once, posing risks to immutability (or lack thereof) [117]. *MethodArgumentCouldBeFinal* (26.31%) and *SystemPrintln* (9.83%) were found next. The former is similar to *LocalVariableCouldBeFinal*, whereas the latter highlights users' penchant to forgo Java's logging functionality and instead redirect outputs to their standard output streams.

Regarding readability, code should adhere to the readability conventions specific to each programming language, and also be readable to minimise confusion encountered by fellow programmers in an effort to minimise technical debt [11]. In SQL, 262,442 of 357,986 parsable snippets violated readability (73.31%). *LT02 (layout.indent)* was the most widespread, covering 15.27% of all violations, followed by *LT01 (layout.spacing)* at 12.99%. Both violations suggest a widespread lack of attention to consistent spacing with respect to the code's visual structure. Additionally, *LT09 (layout.select_targets)* covered 11.23%, suggesting a tendency among users to cram multiple columns into a single line in SELECT statements. This practice may quickly hinder comprehension if the query is complex. 72,858 (21.40%)

JavaScript snippets violated readability conventions. *no-unused-vars* is the most-occurring, apparent in 90.12% of all violations, followed by *no-extra-semi* in 2.72%. Both contribute to code bloat, especially the latter given that JavaScript syntax does not necessitate semicolons at the end of statements. *no-redeclare* covered 2.44% violations, indicating that users are inclined to redeclare variables.

For Python, 157,993 snippets (69.28%) returned at least one readability violation. *E231 (no space around delimiters)* was the most widespread, at 29.46%, implying that users tend to omit spacing between punctuation. *E501 (line too long)* followed suit at 11.25% and *E111 (indentation invalid multiple)* emerged at 8.17%. The former may be attributed to users' tendency to write long Python one-liners, inhibiting the specific elements' identification. The latter indicate inconsistencies in indentation depth, obscuring visual separation.

In Ruby, 97,725 snippets (85.39%) raised at least one violation. *Style/StringLiterals* is the most common, accounting for 13.56% of violations. This observation suggests users tend to use double quotation marks for constructing string literals instead of singles. While both styles are syntactically valid, adhering to a consistent style (i.e., single quotes) enhances comprehensibility. *Style/MethodCallWithArgsParentheses* appeared in 8.22% and *Style/DocumentationMethod* in 6.30%. These findings indicate that users frequently included unnecessary parentheses in method calls and omitted documentation comments for public methods, making complex code harder to understand.

---

[36] https://262.ecma-international.org/5.1

[37] https://0.30000000000000004.com/#ruby

Lastly, 72,169 (73.28%) of Java snippets violated readability. The most recurring violation is *MissingJavadocMethod* (17.45%), indicative of users' lack of attention to include Javadoc comments where necessary. This was then succeeded by *LeftCurly* (14.61%) and *LineLength* (8.62%). These hint that users frequently started code blocks on new lines, yet also tend to compose lines exceeding 100 characters. Extended lines may hinder effective scanning, making the code challenging to understand.

In terms of the performance dimension, snippets should be efficient in performing the intended task with minimal processing steps, in a way that would exhibit efficacy. For SQL, 89,121 (24.90%) returned at least one performance violation. *SELECT ** was the most common one (42.85%) which corroborates prior findings that database programmers tend to favour wildcard selects [118], ostensibly driven by a reluctance to manually type out column names. Following closely behind were *Spaghetti Query Alert* (20.06%) and *UNION Usage* (16.93%). These findings hint at users' preference to write complex, tightly-coupled queries, as well as their lack of awareness regarding the more efficient *UNION ALL* statement, which achieves the same functional outcome as *UNION* without performing duplicate elimination [118].

Interestingly, only a small portion (1,776; 0.52%) of JavaScript snippets raised at least one performance violation. The most notable issue is *no-redeclare* (93.56%), identifying users' tendency to redeclare variables using the *var* keyword. *no-debugger* was observed in 3.44%, indicative of debugger statements within snippets, while *no-fallthrough* appeared in 1.86%. This is indicative that users tend to forgo *break* keywords in *switch-case* statements.

Similarly for Python, a small portion (22,505; 9.87%) of all snippets violated performance rules. *F401 (unused imports)* is the most occurring violation (51.73%) which hints at the prevalence of unused package imports, creating performance overhead. In a similar fashion, *F841 (unused variable)* followed at 17.91%, suggesting that defined variables are mostly left unused. *E402 (import placement*) came next at 13.45%, indicating that users do not place *import* statements at the top of their snippets.

Under Ruby, only 1,322 snippets (1.16%) violated performance rules. The most common problems pertain to the widespread usage of slower methods despite faster alternatives being available. For one, *RegexpMatch* was observed the most (28.68%), suggesting the widespread usage of conventional *match* instead of the faster *match?*. Another example is *StringReplacement* at 14.65%, hinting a widespread usage of *gsub* methods in string literals despite alternatives like *tr* or *delete*. *RedundantBlockCall* followed at 10.34% which implies users' tendency to use *&block* and *block.call* in generators instead of the widely-accepted *yield* keyword.

Regarding Java, out of all 8,450 compilable snippets, 198 (2.34%) returned at least one performance violation. The most occurring violation is *SBSC_USE_STRINGBUFFER_CONCATENATION* (25.89%), implying inefficient string concatenation using loops and the plus operator. Other prevalent violations largely pertain to the lack of *static* modifiers, such as *SIC_INNER_SHOULD_BE_STATIC_ANON* (20.54%), *SS_SHOULD_BE_STATIC* (11.16%), and *SIC_INNER_SHOULD_BE_STATIC* (9.38%).

Finally, security vulnerabilities are defined as codes that can be exploited by malicious third-party attackers. We inspected the degree of security hotspots raised for each language. First, 103,892 (29.02%) of parsable SQL snippets raised at least one violation. Similar to the performance dimension, *SELECT ** was the most frequent (38.22%), followed by *NULL Usage* (37.74%), which is raised when users use *NULL* in comparison expressions. *Spaghetti Query Alert* was also prevalent (17.89%) hinting at the continued usage of complex queries.

For JavaScript, 39,215 snippets (11.52%) had security vulnerabilities. The most frequent issue was *object-injection* (94.08%), hinting at the almost-universal practice of using square bracket notation in objects as operands. The violations *non-literal-regexp* (1.59%) and *non-literal-fs-filename* (1.07%) pertain to dynamic string literals. The former relates to dynamic inputs in regular expressions whereas the latter relates to file paths.

Within Python, 18,025 snippets (7.90%) raised security violations. *S311 (pseudorandom)* stood out as the most prevalent (23.4%), presenting as a stark contrast to JavaScript, given that none of its violations pertain to pseudo-random generators. This finding suggests that Python developers have a higher tendency to utilise dedicated pseudo-random number generators. *S101 (assert usage)* followed closely (14.63%), points to the frequent usage of *assert* statements. *S113 (request without timeout)* appeared in 10.96% of all violations, hinting at the usage of *requests* without timeout mechanisms.

Ruby violations are much less widespread, with only 944 (0.82%) security hotspots. *Open* was the most frequent hotspot (53.94%) highlighting the frequent practice of opening URLs directly with unsanitised inputs. Related to code injection, *Eval* was also frequent (26.15%) which is raised whenever string literals are evaluated as code. *YAMLLoad* followed next (15.48%), flagging the unsafe parsing and loading of YAML objects.

In Java, 107 (1.27% out of all compilable) snippets exhibited security violations. Most violations relate to access modifiers. *MS_SHOULD_BE_FINAL* was the most prevalent (40.12%), and signifies users' penchant to omit *final* non-access modifiers in public static fields. *MC_OVERRIDABLE_METHOD_CALL_IN_CONSTRUCTOR* at 23.46% implies overridable methods are often called within class constructors. *MS_FINAL_PKGPROTECT* was flagged in 7.41% of all instances, again hinting at the omission of *final* and *protected* within mutable public static fields.

## 4.2 Code Snippet Violation Densities (RQ2)

This section presents results for RQ2: *How do coding practices among US contributors differ across states and cities in the United States?*
To delve beyond initial insights, we shift towards a fine-grained examination of code quality variations within US states and cities. While comprehensive city-level results are available in our replication package [27][38], Table 13 highlights that states with established tech hubs (California, New York, Washington, Massachusetts, Texas) consistently produce more total and parsable code snippets in all five studied languages, which may be explained by their proportionately larger pools of users. To enable meaningful comparisons, we divided the total number of snippets for each state by the number of users in that state to get the average number of snippets per user in that particular state, with higher values indicating higher snippet output. Major tech hubs like California and New York consistently produce a high volume of code snippets, yet they do not necessarily exhibit the highest snippet output per user. Instead, these states typically fall within the upper-middle range for each language.

For instance, an average user in California would output 0.903 Python snippets (0.576 parsable), both figures being the 7$^{th}$ highest nationwide. Interestingly, rural states often exhibit either the highest or lowest snippet output nationwide, occupying both extremes. Continuing our Python example, an average user in Alaska would have 1.466 snippets (1.169 parsable), which are the highest figures nationwide. Yet at the same time, users of North Dakota on average contribute 0.234 Python snippets (0.142 parsable), both being the lowest figures across all states. This trend is consistent for other languages as well (refer to our replication package [27] for the full snippet-per-user results[39]). These findings suggest that snippet output is not solely determined by the presence of technology hubs or whether a state is considered more rural or more urban.

To account for potential biases and ensure a fair comparison of code quality across regions, we employed violation density (see Section 3.6). This metric controls for the influence of parsable code volume, preventing regions with more snippets (e.g., California) from appearing to have higher violation rates and thus erroneously suggesting worse code quality.

[38] Replication package » Results » City-Level Parsables.xlsx
[39] Replication package » Results » Snippets per User.xlsx

Table 13. Parsable snippets for each language and state

| State | SQL Snippets (Parsable) | JavaScript Snippets (Parsable) | Python Snippets (Parsable) | Ruby Snippets (Parsable) | Java Snippets (Parsable) |
|---|---|---|---|---|---|
| Alabama | 3,013 (1,045) | 3,149 (1,871) | 1,503 (1,063) | 1,063 (580) | 2,324 (1,450) |
| Alaska | 1,592 (608) | 1,195 (791) | 1,769 (1,411) | 440 (221) | 285 (127) |
| Arizona | 18,426 (10,722) | 9,699 (6,497) | 4,373 (2,765) | 9,886 (7,276) | 4,190 (2,233) |
| Arkansas | 3,709 (1,751) | 2,279 (1,541) | 1,557 (1,056) | 943 (445) | 1,318 (714) |
| California | 165,304 (57,102) | 145,492 (98,173) | 124,984 (79,722) | 76,587 (43,072) | 63,237 (34,870) |
| Colorado | 35,320 (15,601) | 23,230 (15,404) | 13,126 (8,603) | 11,406 (6,473) | 10,029 (5,411) |
| Connecticut | 7,571 (3,987) | 5,403 (3,844) | 4,113 (2,893) | 2,790 (1,566) | 1,526 (787) |
| Delaware | 1,371 (327) | 903 (619) | 660 (457) | 411 (197) | 414 (214) |
| District of Columbia | 29,976 (12,602) | 20,933 (14,554) | 17,502 (11,597) | 11,763 (6,305) | 9,278 (4,882) |
| Florida | 29,509 (13,004) | 26,820 (18,526) | 10,058 (6,752) | 9,285 (5,162) | 8,643 (4,582) |
| Georgia | 22,285 (8,319) | 15,388 (9,920) | 9,277 (6,260) | 7,375 (4,159) | 12,473 (6,652) |
| Hawaii | 1,836 (931) | 1,300 (820) | 2,348 (1,892) | 776 (473) | 402 (212) |
| Idaho | 4,417 (2,019) | 2,150 (1,351) | 1,637 (1,064) | 915 (428) | 1,433 (844) |
| Illinois | 32,378 (15,275) | 23,870 (15,049) | 15,148 (10,360) | 10,250 (5,530) | 9,020 (4,802) |
| Indiana | 7,766 (3,115) | 6,575 (4,405) | 3,631 (2,464) | 3,116 (1,858) | 3,304 (1,648) |
| Iowa | 5,554 (2,578) | 4,066 (2,677) | 1,929 (1,328) | 2,114 (1,505) | 1,409 (717) |
| Kansas | 2,384 (888) | 1,944 (1,340) | 1,176 (847) | 659 (340) | 738 (360) |
| Kentucky | 4,368 (2,117) | 3,229 (2,299) | 1,131 (714) | 1,102 (636) | 1,018 (530) |
| Louisiana | 9,905 (3,078) | 6,921 (4,618) | 4,399 (2,913) | 3,004 (1,435) | 5,180 (3,505) |
| Maine | 2,293 (851) | 2,007 (1,432) | 755 (503) | 422 (182) | 473 (223) |
| Maryland | 13,909 (5,593) | 11,909 (8,059) | 10,038 (7,456) | 3,832 (1,957) | 5,698 (3,173) |
| Massachusetts | 49,563 (16,654) | 36,129 (24,464) | 40,617 (29,631) | 18,258 (9,804) | 17,308 (9,343) |
| Michigan | 15,608 (7,096) | 11,899 (8,219) | 5,409 (3,466) | 4,434 (2,405) | 5,005 (2,785) |
| Minnesota | 15,024 (7,141) | 11,466 (7,744) | 4,863 (2,994) | 4,667 (2,549) | 4,780 (2,828) |
| Mississippi | 985 (444) | 940 (591) | 524 (342) | 193 (61) | 375 (214) |
| Missouri | 20,063 (10,869) | 8,066 (5,329) | 4,997 (3,321) | 4,066 (2,092) | 4,269 (2,324) |
| Montana | 927 (297) | 1,009 (623) | 506 (319) | 327 (130) | 350 (207) |
| Nebraska | 2,830 (1,005) | 3,787 (2,757) | 681 (454) | 1,095 (704) | 941 (470) |
| Nevada | 4,445 (1,265) | 3,497 (2,232) | 1,937 (1,371) | 2,160 (1,323) | 1,161 (528) |
| New Hampshire | 2,893 (1,202) | 2,197 (1,545) | 794 (492) | 895 (439) | 941 (503) |
| New Jersey | 17,897 (7,188) | 12,286 (8,227) | 7,862 (5,282) | 3,201 (1,372) | 5,654 (3,002) |
| New Mexico | 5,123 (801) | 2,274 (1,099) | 1,575 (983) | 1,380 (676) | 1,017 (440) |
| New York | 172,753 (117,263) | 69,159 (47,345) | 51,391 (33,739) | 36,746 (20,724) | 25,700 (13,670) |
| North Carolina | 33,665 (12,753) | 33,345 (25,216) | 12,972 (8,784) | 10,368 (5,407) | 11,595 (6,470) |
| North Dakota | 363 (145) | 421 (279) | 181 (110) | 216 (122) | 163 (84) |
| Ohio | 20,162 (8,378) | 18,295 (13,195) | 7,175 (4,936) | 5,409 (2,965) | 6,182 (3,426) |
| Oklahoma | 4,023 (1,496) | 3,826 (2,687) | 5,171 (4,208) | 1,297 (628) | 1,158 (635) |
| Oregon | 32,290 (15,350) | 21,918 (14,992) | 12,855 (8,740) | 11,719 (6,418) | 7,914 (4,152) |
| Pennsylvania | 22,459 (8,260) | 20,557 (13,720) | 13,438 (8,649) | 8,684 (4,943) | 11,892 (6,852) |
| Rhode Island | 4,711 (3,663) | 1,376 (958) | 674 (410) | 502 (270) | 384 (215) |
| South Carolina | 6,210 (2,306) | 5,932 (4,135) | 2,706 (1,716) | 1,685 (914) | 6,569 (4,900) |
| South Dakota | 1,012 (500) | 870 (671) | 233 (161) | 154 (51) | 365 (221) |
| Tennessee | 8,346 (3,337) | 9,755 (6,858) | 3,277 (2,144) | 3,056 (1,775) | 2,688 (1,459) |
| Texas | 76,958 (27,485) | 63,533 (43,091) | 35,244 (23,995) | 24,589 (13,226) | 25,840 (14,090) |
| Utah | 10,224 (3,848) | 10,960 (7,468) | 6,716 (4,614) | 3,816 (2,148) | 3,497 (1,859) |
| Vermont | 1,841 (754) | 1,308 (845) | 1,014 (695) | 1,237 (800) | 366 (162) |
| Virginia | 14,351 (5,834) | 11,186 (7,484) | 6,633 (4,143) | 4,782 (2,580) | 5,858 (3,179) |
| Washington | 53,827 (18,460) | 41,497 (26,536) | 40,039 (25,313) | 19,019 (10,299) | 21,822 (11,628) |
| West Virginia | 5,573 (2,286) | 3,638 (2,266) | 2,349 (1,638) | 1,946 (904) | 2,046 (1,115) |
| Wisconsin | 10,014 (3,381) | 12,196 (8,262) | 3,238 (2,231) | 3,109 (1,659) | 3,763 (1,967) |
| Wyoming | 672 (228) | 512 (310) | 347 (224) | 368 (233) | 198 (109) |

For the reliability dimension, New Mexico held the highest density for SQL (0.101) and JavaScript (0.576), suggesting concerns regarding error-proneness for its users. Similarly, Mississippi had the highest violation density for Ruby (0.374) and Java (0.325), while Missouri led in Python violations (0.482). Counter to what might be intuitively expected, major tech hubs like California and New York were concentrated in the middle of the violation density spectrum, while smaller tech states had lower densities. These findings suggest that factors beyond tech industry presence influence code reliability, which is explored in Section 5.2. The pattern of moderate tech ecosystems having higher violation densities extends to the city level. Champaign, IL exemplifies this, with the highest observed JavaScript reliability violation density (0.592). Interestingly, even major tech hubs like San Francisco, CA and Seattle, WA do not necessarily have the highest or lowest violations. For one, San Francisco's reliability violation density for Python (0.328) is clustered on the median (43rd highest), and for Ruby (0.112) is lower than most cities (75th).

Looking at the readability dimension reveals a surprising trend. Unlike reliability, states with less developed tech sectors (Alabama, Maine, North Dakota, Mississippi) exhibited high violation densities (e.g., Alabama's highest SQL density at 0.772). These findings suggest a relationship between tech maturity of a state and developer code clarity. However, increasing technological influence may not offer a panacea, as prominent tech hubs (California, New York) also had significant readability violations (e.g., Massachusetts's Python density of 0.386, 15th highest). Conversely, states with moderate tech ecosystems had the lowest readability violations (South Carolina's lowest Java density at 0.081). This suggests a potential "sweet spot" where established practices promote clarity without the complexities of advanced ecosystems. Similar patterns emerge at the city level, with Davis, CA (large suburban area) having a low Java violation density (0.056). This reinforces the notion that a balanced level of technological integration might be optimal for code clarity, avoiding both the pitfalls of nascent and highly advanced ecosystems. However, deviations exist, like Lansing, MI (state capital) having the lowest Ruby density (0.545). These findings highlight the multifaceted nature of factors influencing code readability, beyond just technological maturity. In states with advanced tech sectors, for example, a diverse range of developers with varying levels of experience may be attracted. This diversity could lead to a balancing effect on code readability, preventing extreme deviations from the average level and resulting in a more moderate level of readability violations (explored further in Section 5.2).

Analysis of performance violation density reveals that states with established tech sectors (California, New York, Massachusetts) have lower densities across languages (e.g., New York's SQL density of 0.017). These findings suggest the demand for performant code within these competitive, large, and vibrant tech landscapes (see Section 5.2). Conversely, states with less developed tech sectors (West Virginia, Vermont, and North Dakota) have higher densities (West Virginia's highest SQL at 0.178). However, this trend does not hold for all languages, exemplified by Maryland with lowest JavaScript density of 0.003. This suggests other factors beyond tech maturity influence performance violations, like regional practices (e.g., Washington's low Ruby density of 0.006). Interestingly, city-level data shows an inverse relationship — larger cities that are not necessarily tech hubs have higher densities. For instance, Irving, TX has 0.271 for SQL. These discoveries accentuate the presence of less nuanced, multifaceted factors which determines performant coding practices.

Like performance, security violation density analysis suggests a link to urbanisation. States with larger urban populations (California, New Jersey, and New York) have lower densities. Examples include New York's density of 0.0015 and New Jersey's 0.0017 for Python. However, the trends were less stark than the performance dimension, in which several deviations still occurred. Examples include Nevada's high Java violation density (0.233) and Florida's high Ruby density (0.0086), highlighting the non-monotonic relationship between the urbanisation of a state and their secure coding practices. An inverse trend emerged on rural states, exhibiting higher violation densities. North Dakota and Maine, for instance, boasted the highest violation densities for JavaScript at 0.037 and 0.036, respectively. These findings can also be seen to ripple towards city-level despite some fluctuations. Bustling metropolises generally exhibited lower violation densities across various programming languages. Boca Raton, FL, has one of the lowest densities for Java (0.055) and Houston, TX exhibiting the same notion for Python (0.004). In general, our analysis reveals a potential causal relationship between violation densities across all dimensions and regional urbanisation.

### 4.3 Code Snippet Relations (RQ3)

This section presents results for RQ3: *How do diversity indicators of US regions affect code quality?* Our preliminary findings indicate that certain census diversity indicators exhibit moderate to strong negative correlations with the states' violation densities. Negative correlation infers that an increase of that indicator would lead to a decrease in violation density (i.e., better quality). Results showed all pairs' $p$-values to be less than 0.05, confirming that all correlation pairs are unlikely due to chance. In subsequent reporting of the results, only medium ($r$ = 0.3–0.5) to strong correlations ($r \geq 0.5$) (as per Cohen [119]) were highlighted, allowing examination and comparison of cross-region diversity and code quality densities. Firstly, the indicator *Gini index* was shown to moderately correlate with JavaScript Performance ($r$ = -0.392), as well as Ruby Readability (-0.3) and Reliability (-0.323). This implies that states with higher *Gini index*, which signals better wealth equality, tend to also yield less violations. We also found the indicator *households with one or more types of computing devices* exhibited strong negative correlations with SQL Performance (-0.597) and Ruby Readability (-0.527), and also moderate negative correlations with multiple violation densities including Python Reliability (-0.457) and Performance (-0.363), as well as SQL Security (-0.455) and Readability (-0.383). These results hint that the prevalence of domestic information and communications technology (ICT) access at the state-level tends to positively impact code quality. Another ICT-related indicator that is worth noting is *Internet Subscriptions in Household: With an Internet subscription*, exhibiting strong negative correlations with Java (-0.659) and Python Security (-0.652). It also presented moderate negative correlations against Ruby Security (-0.447), Python Readability (-0.443) and Reliability (-0.348), Java Performance (-0.406) and Readability (-0.395), as well as JavaScript Performance (-0.315) and Readability (-0.306). This supports the idea that domestic ICT access can bolster code quality regardless of dimensions and languages.

Apart from ICT-related indicators, *per capita income in the past 12 months* negatively correlates with Java (-0.578) and Python Security (-0.408), i.e., higher regional income is correlated with more secure coding practices. The racial diversity indicator *white alone (% of all population)* positively correlates with Java Readability (0.314), i.e., lower racial diversity in a state correlates with lower code readability. Other sets of positive correlations were found between *total unemployment* and Python (0.358) and SQL Readability (0.417), i.e., higher unemployment correlates with less readable code. In terms of education, *school enrolment by level of school for the population 3 years and over: enrolled up to grade 12* negatively correlates with three performance violation densities: Java (-0.45), JavaScript (-0.404), and Python (-0.318). That is, broader coverage of compulsory basic education correlates with more performant code. Finally, *Total Females* was found to correlate with Ruby Readability (-0.653) and Security (-0.564), JavaScript Security (-0.585) and Readability (-0.524), Python Performance (-0.49) and Readability (-0.43), and Java Security (-0.409) and Performance (-0.321); while *Total Males* was found to correlate with: Ruby Security (-0.616), Python Reliability (-0.536) and Security (-0.415), Java Reliability (-0.469) and Security (-0.48), and SQL Reliability (-0.385). These findings may hint at a relationship between gender composition and code quality metrics, but further research is clearly needed to understand these dynamics. Table 14 below documents the strong and moderate correlations for all census indicators and code quality violation densities, while our replication package hosts the full correlation matrix[40] [27].

Table 14. State-level indicators' correlations

| Diversity Indicator | Code Quality Violation Density | Pearson's $r$ |
|---|---|---|
| *Gini index* | Ruby Reliability | -0.323 |
| | Ruby Readability | -0.300 |
| | JavaScript Performance | -0.392 |
| *Households with one or more types of computing devices* | SQL Security | -0.455 |
| | SQL Readability | -0.383 |
| | SQL Performance | -0.597 |
| | Ruby Readability | -0.527 |
| | Ruby Performance | -0.492 |
| | Python Reliability | -0.457 |
| | Python Performance | -0.363 |
| | JavaScript Reliability | -0.302 |

[40] Replication package » Results » Census Indicators Correlations (RQ3) » Census Indicators Correlations.xlsx

| Diversity Indicator | Code Quality Violation Density | Pearson's *r* |
|---|---|---|
| | Java Performance | -0.327 |
| *Internet Subscriptions in Household: With an Internet subscription* | Ruby Security | -0.447 |
| | Python Security | -0.652 |
| | Python Reliability | -0.348 |
| | Python Readability | -0.443 |
| | JavaScript Readability | -0.306 |
| | JavaScript Performance | -0.315 |
| | Java Security | -0.659 |
| | Java Readability | -0.395 |
| | Java Performance | -0.406 |
| *Per capita income in the past 12 months (in 2020 inflation-adjusted dollars)* | Python Security | -0.408 |
| | Java Security | -0.578 |
| *Race: White alone (% of all population)* | Java Readability | 0.314 |
| *School Enrolment by Level of School for the Population 3 Years and Over: Enrolled up to Grade 12* | Python Performance | -0.318 |
| | JavaScript Performance | -0.404 |
| | Java Performance | -0.450 |
| *Total Females* | Ruby Security | -0.564 |
| | Ruby Readability | -0.653 |
| | Python Readability | -0.430 |
| | Python Performance | -0.490 |
| | JavaScript Security | -0.585 |
| | JavaScript Readability | -0.524 |
| | Java Security | -0.409 |
| | Java Performance | -0.321 |
| *Total Males* | SQL Reliability | -0.385 |
| | Ruby Security | -0.616 |
| | Python Security | -0.415 |
| | Python Reliability | -0.536 |
| | Java Security | -0.480 |
| | Java Reliability | -0.469 |
| *Total unemployment* | SQL Readability | 0.417 |
| | Python Readability | 0.358 |

## 4.4 Code Snippet Themes (RQ4)

This section presents results for RQ4: *What themes are prevalent in code quality violations, and how do these themes differ across US regions?*
Following our inductive content analysis, results were aggregated using frequency counts and percentages. Snippets with more than one violation were assigned multiple coding themes. Table 15 summarises the emergent coding themes discovered across all quality dimensions, their descriptions, and chi-squared test results. Significance levels are appended at the end of each test statistic following the 'three-star system' (i.e., *$p \leq 0.05$, **$p \leq 0.01$, ***$p \leq 0.001$) [120]. To further illustrate discovered trends, Figure 7 provides a visual representation of the quantitative analysis results for all three states across each quality dimension. Detailed examples and analytical merit of each coding theme may otherwise be seen in our replication package [27][41].

Firstly, California exhibited the highest prevalence of complex non-basic reliability errors. These errors, exemplified by *Unclosed Resource* (34 occurrences, 6.79%) are notably more frequent in California compared to Utah (25 occurrences, 3.86%) and North Dakota (14 occurrences, 2.08%). Several other coding themes that adhere to this trend are *Race Conditions* (45; 8.98%), and *Field Mutability* (40; 7.98%) in California, which is the highest of all three states. In contrast, somewhat basic reliability violations such as *Incorrect Branching* (10 times; 7.30%) and *Incorrect Loops* (8 times; 5.84%) were the least frequent in California. These findings suggest that a state's tech workforce and R&D presence may impact the complexity of their developers' reliability violations. However, certain deviations still exist, such as the prevalence of *Usage Mismatch* violations (68 occurrences, 13.57%) in California compared to Utah (68 occurrences, 10.49%), and North Dakota (29, 4.31%).

[41] Replication package » Coding Themes » Reliability to Security

Table 15. Discovered content analysis coding themes

| Dimension | Scale | Coding Theme | Description | $\chi^2$ |
|---|---|---|---|---|
| Reliability | 1 | Field Mutability | Fields that are only used once are not given immutable access modifiers. | 2.268 (ns) |
| | 2 | Built-In Mutability | Immutable built-in data types are modified. | 10.634** |
| | 3 | Excessive Imports | Numerous imported packages that remain unutilised, making the code brittle. | 2.032 (ns) |
| | 4 | Usage Mismatch | Non-built-in methods are used in a way that does not comply with its intended use-case. | 21.509*** |
| | 5 | Redundant Statements | Unnecessary statements that risk accidental field misuse or modification. | 11.594** |
| | 6 | Empty Exceptions | Exceptions are caught yet are ignored and unhandled. | 27.908*** |
| | 7 | Overly-Broad Exceptions | General, non-explanatory raw exception types. | 14.851*** |
| | 8 | Lack of Interfaces | Insufficient abstractions and interfaces, hindering versatility and extensibility. | 8.109* |
| | 9 | God Class | Classes that contain too many responsibilities and functionalities. | 6.591* |
| | 10 | Syntax Errors | General syntax errors that will prevent the code from running. | 19.289*** |
| | 11 | Threading Errors | Errors that are related to multithreading and will result in deadlocks. | 18.697*** |
| | 12 | Data Structures | Improper utilisation for data structures' functionalities | 0.327 (ns) |
| | 13 | Built-in Methods | Built-in methods are used in a way that deviates from its original design. | 44.189*** |
| | 14 | Incorrect Loops | Loops that end prematurely or goes on infinitely. | 24.56*** |
| | 15 | Incorrect Branching | Program flow may deviate from the expected behaviour. | 18.221*** |
| | 16 | Race Conditions | Multiple threads process a shared item where the final result depends on the order of executions. | 17.345*** |
| | 17 | Unclosed Resource | Connections and resources that are not closed properly following its utilisation. | 8.803* |
| | 18 | Not Coded | — | 0.779 (ns) |
| Readability | 1 | Implicit Syntax | Non-self-explanatory syntaxes and expressions. | 7.022* |
| | 2 | Advanced Language Features | Language-specific features that are unintuitive. | 15.145*** |
| | 3 | Complicated Methods | Complex methods despite easier alternatives can be implemented. | 20.086*** |
| | 4 | Long one-liners | Lines that excessively chain instructions, hindering interpretability. | 6.976* |
| | 5 | Separable Statements | Lines that contain several statements that would be more interpretable if separated. | 2.439 (ns) |
| | 6 | Redundant Stylistic Choice | Stylistic choices that are unnecessary and does not affect program behaviour. | 9.909** |
| | 7 | Irregular Spacing | Inconsistent whitespaces and indentations between array elements, code blocks, and statements. | 3.006 (ns) |
| | 8 | Irregular Comments | Comments that are inconsistent or misleading. | 9.052* |
| | 9 | Irregular Empty Lines | Too much or too little empty lines between statements. | 12.566** |

| Dimension | Scale | Coding Theme | Description | $\chi^2$ |
|---|---|---|---|---|
| | 10 | Lack of Documentation | Methods, functions, or classes that are undocumented. | 7.965* |
| | 11 | Incomplete Documentation | Documentations that are vague, unclear, and/or hardly understandable. | 0.058 (ns) |
| | 12 | Single-Letter Names | Object or variable names that only contain a single letter. | 5.115 (ns) |
| | 13 | Ambiguous Names | Non-self-explanatory object or variable names. | 15.542*** |
| | 14 | Not Coded | — | 4.511 (ns) |
| Performance | 1 | Small Memory Allocation | Minor performance overheads that only slightly affects performance. | 22.509*** |
| | 2 | Invalid Regular Expressions | Regular expressions that does not return any matches. | 3.813 (ns) |
| | 3 | Invalid Character Classes | Multiple code point characters in character class syntax. | 15.71*** |
| | 4 | Unnecessary Computation | Usage of complex algorithmic steps where a faster alternative may be implemented. | 73.349*** |
| | 5 | No Lazy Computation | Lack of lazy implementations where computations are not done ad-hoc. | 14.962*** |
| | 6 | Comparisons to Sort | Usage of comparison operators to sort values. | 27.862*** |
| | 7 | Usage of Inefficient Methods | Usage of built-in language features and methods that are slower than others. | 10.553** |
| | 8 | Fall-through | Switch-case statements without breaking each case. | 3.863 (ns) |
| | 9 | Not Coded | — | 0.839 (ns) |
| Security | 1 | Dynamic Paths | Accession to non-hard-coded file paths. | 18.269*** |
| | 2 | File Parsing | Attempts at parsing unsanitised files such as XML. | 8.673* |
| | 3 | Dynamic Input | Parsing non-literal and non-deterministic inputs. | 25.181*** |
| | 4 | Unsafe Shell Injection | Usage of unsafe shell methods. | 1.598 (ns) |
| | 5 | Code Evaluation | Parsing arbitrary strings and evaluating them as code statements. | 2.763 (ns) |
| | 6 | Unsafe Deserialisation | Deserialising external arbitrary files without prior validation nor sanitisation. | 32.229*** |
| | 7 | Readable Passwords | Passwords stored as plaintext or is decipherable through reverse encoding. | 11.879** |
| | 8 | Silent Errors | Program may not behave unexpectedly in a way that security vulnerabilities are opened. | 0.214 (ns) |
| | 9 | Pseudorandom Numbers | Usage of pseudo-random number generators instead of true random sequences. | 5.826 (ns) |
| | 10 | Arbitrary File Opening | Exposure to insecure APIs or external functionalities that permits arbitrary local file opening. | 5.096 (ns) |
| | 11 | Not Coded | — | 4.904 (ns) |

*$p \leq 0.05$; **$p \leq 0.01$; ***$p \leq 0.001$; (ns) = not significant.

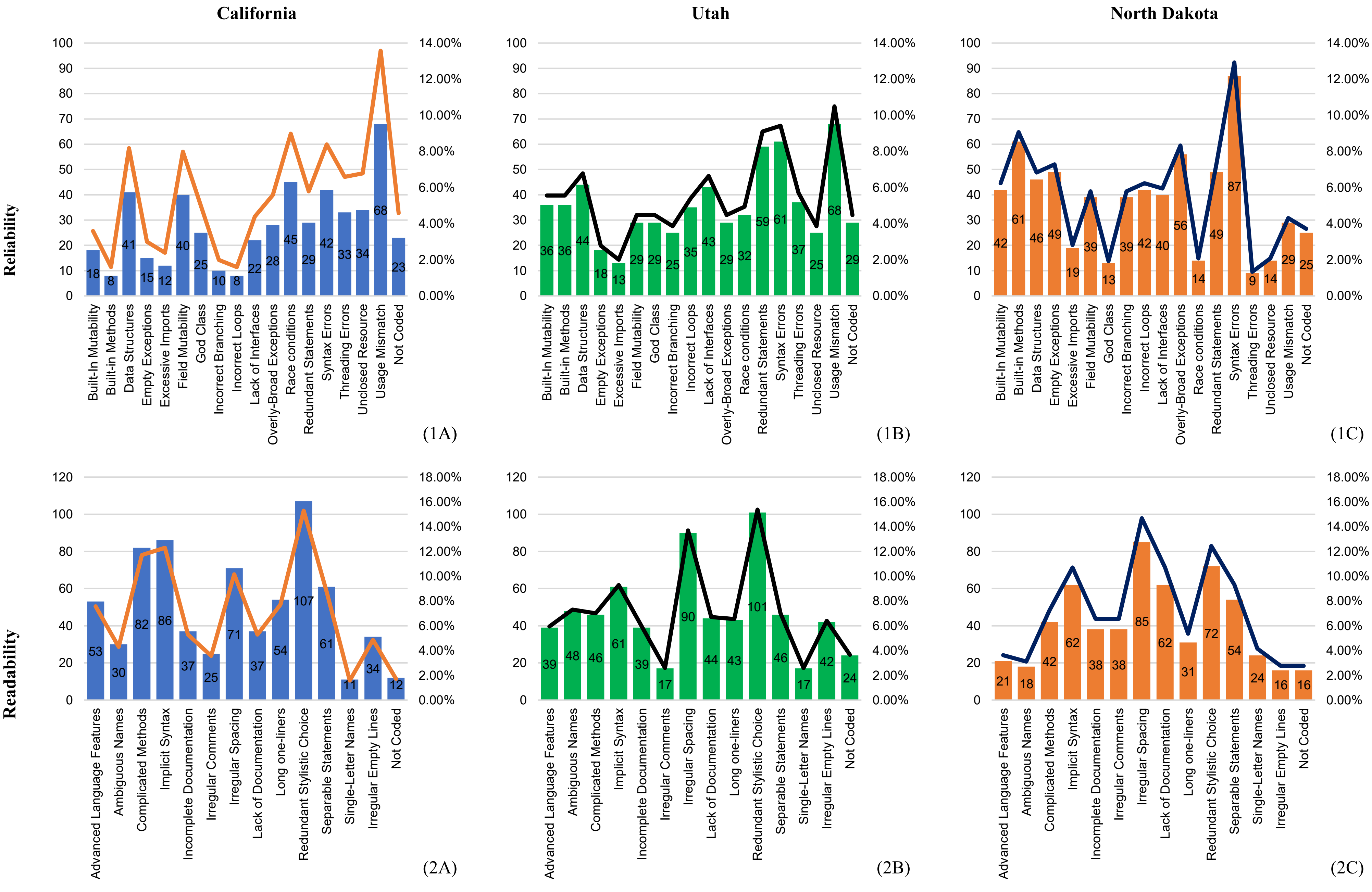
California
Utah
North Dakota
Reliability
Readability
100
90
80
70
60
50
40
30
20
10
0
14.00%
12.00%
10.00%
8.00%
6.00%
4.00%
2.00%
0.00%
18 8 41 15 12 40 25 10 8 22 28 45 29 42 33 34 68 23
36 36 44 18 13 29 29 25 35 43 29 32 59 61 37 25 68 29
42 61 46 49 19 39 13 39 42 40 56 14 49 87 9 14 29 25
Built-In Mutability
Built-in Methods
Data Structures
Empty Exceptions
Excessive Imports
Field Mutability
God Class
Incorrect Branching
Incorrect Loops
Lack of Interfaces
Overly-Broad Exceptions
Race conditions
Redundant Statements
Syntax Errors
Threading Errors
Unclosed Resource
Usage Mismatch
Not Coded
(1A)
(1B)
(1C)
120
100
80
60
40
20
0
18.00%
16.00%
14.00%
12.00%
10.00%
8.00%
6.00%
4.00%
2.00%
0.00%
53 30 82 86 37 25 71 37 54 107 61 11 34 12
39 48 46 61 39 17 90 44 43 101 46 17 42 24
21 18 42 62 38 38 85 62 31 72 54 24 16 16
Advanced Language Features
Ambiguous Names
Complicated Methods
Implicit Syntax
Incomplete Documentation
Irregular Comments
Irregular Spacing
Lack of Documentation
Long one-liners
Redundant Stylistic Choice
Separable Statements
Single-Letter Names
Irregular Empty Lines
Not Coded
(2A)
(2B)
(2C)

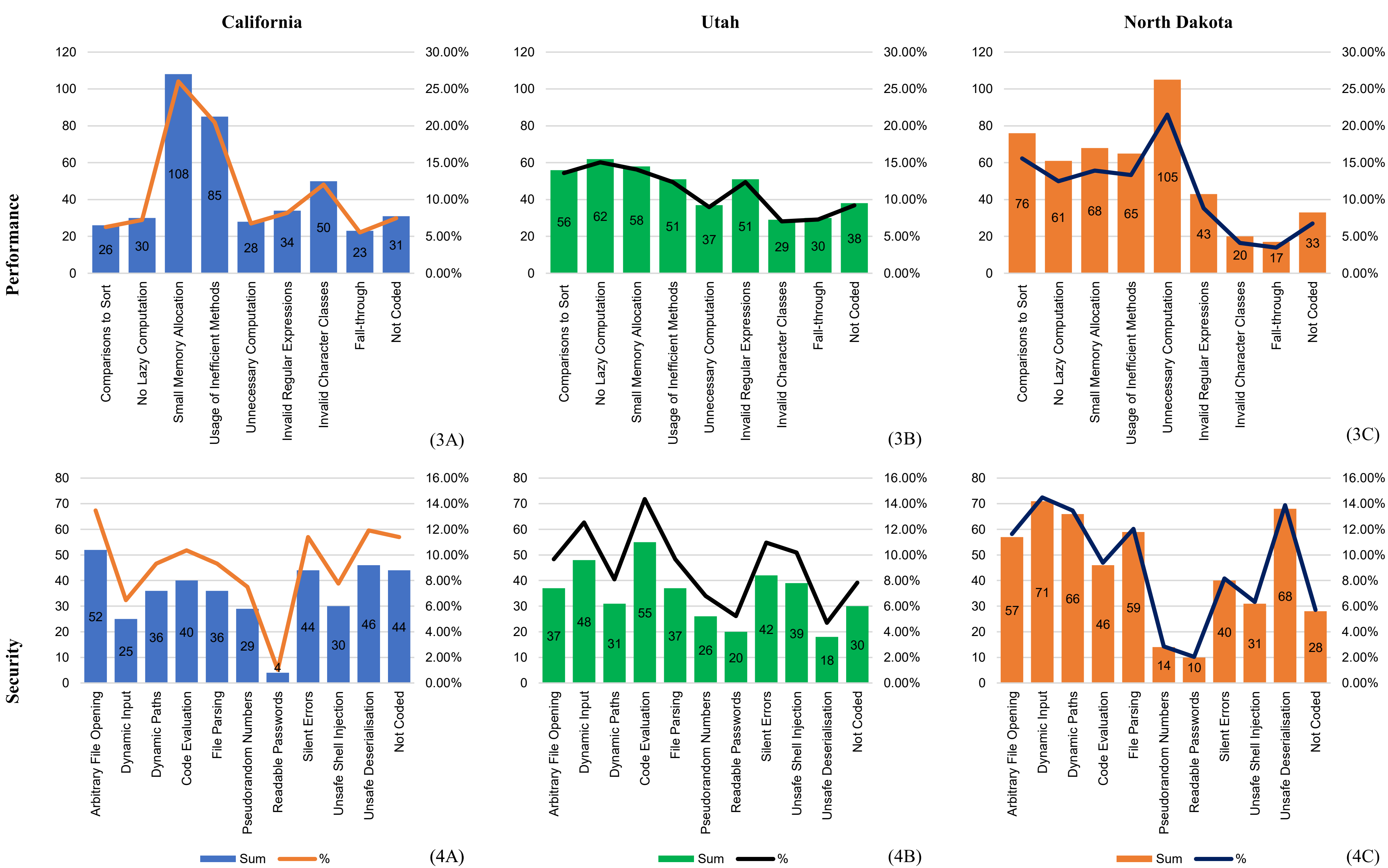

Figure 7. Content analysis results per dimension and state

On the other hand, North Dakota exhibited a tendency towards simpler reliability violations. This is evident in the high frequency of *Syntax Errors* (87 times; 12.93%) compared to Utah (61; 9.41%) and California (42; 8.38%), which prevent a program from running and thus are relatively basic mistakes. Similarly, North Dakota had a higher proportion of violations related to *Overly-Broad Exceptions* (56 times; 8.32%) and *Built-In Methods* (61 times; 9.06%) compared to the other states. These findings further suggest that a region's technological advancement can influence the reliability of code contributions. North Dakota, with a more nascent tech industry, appears to have a higher prevalence of these simpler reliability violations. When examining readability, California users exhibited a tendency to employ more advanced language features, doing so 53 times (7.57%), compared to Utah (39 times; 5.94%) and North Dakota (21; 3.63%). This preference for less intuitive code in California is further supported by the *Implicit Syntax* coding theme, being raised most frequently in California (86 times; 12.29%) compared to the other states. In general, these findings suggest that California users tend to contribute code that is more complex and less explicit, making it less intuitive for beginners. Regarding stylistic choices, both California (107; 15.29%) and Utah (101; 15.37%) exhibited a high prevalence of the *Redundant Stylistic Choice* violation. This finding suggests that developers in these states may deviate more frequently from established coding conventions, employing a wider variety of stylistic approaches.

Certain trends were consistent across all three states, however, suggesting that user behaviour may transcend geographical boundaries. For one, all states exhibited a high frequency of the *Irregular Spacing* violation. Utah and North Dakota were found to have the most instances, with 90 (13.70%) and 85 (14.68%), respectively. These results imply that consistent spacing practices are generally overlooked regardless of where users reside. An analysis of performance-related coding violations revealed that Utah's violations consistently fell below 60 across all coding themes, which contrasts with California and North Dakota where violations for one theme could surpass 100 occurrences. In fact, the most frequent violation in Utah, *No Lazy Computation*, only occurred 62 times (15.05%). Initial findings suggest a generally higher adherence to performant coding practices in Utah. On the other hand, the most frequent violations in California were *Small Memory Allocation* (108 times; 26.02%) and *Usage of Inefficient Methods* (85 times; 20.48%). This suggests that users in California might be more prone to contribute snippets with slight performance overheads that would be detrimental if they accumulate, or that use less efficient methods when faster alternatives exist. Similarly in North Dakota, *Unnecessary Computation* occurred the most often at 105 times (21.52%) which hints that their users tend to contribute unnecessarily complex implementations despite the possibility of a faster alternative. Overall, though, *Invalid Regular Expressions* and *Fall-through* were consistently low across all three states. California had the highest frequency of the former at 34 occurrences (8.19%), while North Dakota had the most *Fall-through* themes at 17 instances (3.48%). These low figures imply that developers in all three states generally handle regular expressions and switch-case constructs effectively.

Finally, in terms of security, California exhibited a relatively low number of security violations overall. However, two themes, *Arbitrary File Opening* (52 times; 13.47%) and *Silent Errors* (44 times; 11.40%), were more frequent in California compared to others. Interestingly, *Arbitrary File Opening* was also prevalent in North Dakota (57 times; 11.63%) but less frequent in Utah (37 times; 9.66%). These findings suggest that users in California and North Dakota might be more likely to provide answers with connection-related mistakes that could accidentally open files. The prevalence of the *Readable Passwords* coding theme served as another indicator of shared practices between California and North Dakota users. Such behaviours were exceptionally rare in both states, occurring only 4 times (1.04%) in the former and 10 times (2.04%) in the latter, suggesting that users in these regions are likely aware of the security pitfalls associated with storing passwords in plain text or unsalted hashes. Finally, snippets from Utah signalled concerning trends wherein *Dynamic Input* (48 occurrences; 12.53%) and *Code Evaluation* (55; 14.36%) were among the most frequent coding themes in that state. These findings hint that users of Utah might be less likely to sanitise user input and have a tendency to evaluate strings as code, which can introduce vulnerabilities to injection attacks.

# 5 DISCUSSION AND IMPLICATIONS

## 5.1 Code Snippet Quality (RQ1)

In this section, we discuss snippet quality with respect to RQ1: *What is the quality of code snippets across different programming languages on Stack Overflow?*

Our analysis revealed a relatively low prevalence of reliability violations. In SQL (2.81%), JavaScript (0.52%), and Python (4.49%) we saw only a small fraction of parsable snippets are prone to errors and bugs. Reliability violations were more frequent in Ruby (21.23%), but Java exhibited the highest rate (84.36%). These findings suggest that users may have an easier time writing more reliable snippets in SQL, JavaScript, and Python compared to Ruby and Java. This difference might be due to Java's complexity and its strict type checking, which can make debugging trickier [24, 83]. As a result, ensuring code reliability in Java might be more challenging. Looking at the violations, *unmatched-parentheses* are the most common error in SQL code snippets shared online. This prevalence likely arises from the context-specific nature of SQL queries. Unlike code intended for independent execution, SQL queries necessitate database access to test its correctness, making pre-testing impractical. We posit this reliance on external execution might lead to a higher frequency of basic syntax errors like unpaired parentheses, compared to more complex errors. Holzmann [121] found that large software repositories were found to harbour more minor errors than major ones, and based on our results, this pattern can also be seen in SQL snippets. In JavaScript, misuse of prototypes and omitting variable definitions are frequent, suggesting a lack of understanding of the language's core prototype concepts. Python shares the issue of missing variable declarations, corroborating Reid et al. [122] where snippets were often found to omit variable declarations, as users are expected to manually check such omissions before using them. Ruby suffers from unutilised code remnants, hinting at residuals of incomplete refactoring [123]. Java developers tend to forgo declaring constants, suggesting a tendency to waive immutability and hence increase the risk of introducing bugs [124]. We also reveal a spectrum of error severity across languages. Some violations, like missing conditions in SQL or type conversions in Ruby, might not cause immediate errors but introduce potential for unexpected behaviour. Others, like undeclared variables in Python or invalid assignments in JavaScript, will halt program execution. Finally, specific violations like improper error handling practices (e.g., using bare except in Python) highlight the need for robust code to ensure resilience against failures.

Under the readability dimension, Ruby suffers the most readability violations (85.39%), followed by SQL (73.31%), Java (73.28%), and Python (69.28%). This might be due to less strict adherence to community conventions in these languages, especially for Python and SQL, which are known for readability [125, 126]. The inherent readability of these languages might lead users to subconsciously overlook established conventions, assuming the code's syntactic clarity is sufficient. Moreover, high readability violations can also be attributed to different coding conventions within each language. For example, SQL has various dialects beyond the ISO standard, such as Oracle Native or BigQuery. Compared to reliability, readability violations were noticeably higher, suggesting a prioritisation of code that works over code that is understandable. Our results corroborate Meldrum et al. [11] in highlighting a potential disregard for readability conventions within the online coding community. The high readability violations might also be partially due to the nature of answer snippets themselves, which tend to be short and focussed on solving a specific problem [11, 85]. Unlike code written for independent use, readability might be less of a concern on Stack Overflow where users prioritise finding solutions quickly [8], leading to less emphasis on readability conventions. The analysis of language-specific violations reveals that improper indentation is the most common issue for SQL, which likely stems from a focus on quick solutions for one-time use cases (i.e., ad-hoc queries) where readability might be less important [127, 128]. JavaScript exhibits frequent violations regarding unused variables. While this could be a stylistic choice, it can disrupt code flow and reduce readability [72, 129], especially when variables are declared in one snippet and used in another within the same post. Readers may experience loss of continuity due to context switching. Figure 8 depicts such an example where *this.var* is first declared in the first (upper) snippet but not used until the second.

In Python, frequent omission of spacing around delimiters in array-like data structures might be due to a perceived lack of impact on functionality, despite going against readability best practices as specified in the PEP 8 style guide [54, 130]. Similarly, Ruby's *Style/StringLiterals* violation, while seemingly minor, reflects a preference for single quotes to avoid unintended variable interpolation within strings using double quotes [131]. For example, *"Hello, #{name}!"* would insert the value of the *name* variable into the string. In contrast, single quotation marks do not present as much versatility, preventing accidental variable substitution and imparting a clearer semantic context to readers. Java violations mainly involve lack of Javadoc comments, which may be due to code snippets being miniscule (as noted above) and therefore contextual information surrounding the snippet should adequately elucidate its functionality [11].

1

Well when you use this in some function `this` is referencing to the object which actually the function is. Here:

```
function myF() {
    this.var = 'hey';
}
```

You can reach `var` using this (myF as a constructor function):

```
var obj = new myF();
alert(obj.var);
```

Figure 8. Example post (Posts.Id 8566320) where variable is declared and used in different snippets

Overall, violations are mostly focussed in practices that would create technical debt by making code harder to understand in the long run. This includes enforcing consistent capitalisation (*capitalisation.keywords* in SQL), avoiding abbreviations (*AbbreviationAsWordInName* in Java), using escape characters properly (*no-useless-escape* in JavaScript), and maintaining clear visual separation between code elements (*LT05* in Python) [132].

For performance, SQL (24.90%) has the most violations, followed by Python (9.87%), Java (2.34%), Ruby (1.16%), and JavaScript (0.52%). These findings suggest users struggle more with optimising SQL queries compared to other languages, potentially owing to the inherent complexity of database operations [133] and of the SQL language itself. This aligns with Lyu et al. [134], where they identified the persistence of SQL performance anti-patterns even in performance-critical applications, significantly impacting both runtime and energy consumption. For instance, using *SELECT ** was shown to consume ten times more energy on average and hinder scalability because it retrieves excessive data [134]. Notably, wildcard queries were violated most often. Behaviour-wise, wildcard querying may be attributed to an unwillingness to manually type column names [118]. JavaScript exhibits frequent variable redeclarations (*no-redeclare*), a practice that can degrade performance over time, even though it might seem like a shortcut initially [135]. In Python, unnecessary imports are somewhat common, especially for standard libraries like *sys* or *os*. This is likely because developers are more familiar with these libraries and might import them "just in case," whereas third-party libraries with specific purposes are less likely to be imported and then not used [156]. Ruby developers frequently use basic *RegexpMatch* statements for regular expressions despite faster alternatives existing [136]. This suggests a lack of awareness, or a preference for more familiar methods instead of novel ones. Similarly, Java developers often resort to the plus operator (+) for string concatenation (*SBSC_USE_STRINGBUFFER_CONCATENATION*) instead of more efficient built-in functions [137]. This again might be due to unfamiliarity, or to a misunderstanding of Java's immutable strings, where concatenation creates new objects and potentially consumes more memory [11]. From a contextual standpoint, performance violations vary in severity. Less critical examples include variable redeclaration (*no-redeclare*) in JavaScript and import placement (*E402*) in Python. Conversely, more serious violations involve inefficient regular expressions (*no-invalid-regexp*) in JavaScript and manual memory management (*DM_GC*) in Java [11, 138]. These findings highlight the importance of code optimisation even in short snippets, as seemingly minor design choices can accumulate and impact performance.

Under the security dimension, SQL queries exhibit the most violations (29.02%) compared to JavaScript (11.52%) and Python (7.90%), while Ruby (0.82%) and Java (1.27%) exhibited lower percentages. The higher frequency of security violations in SQL snippets may partially explain the prevalence of SQL injection attacks observed in the broader SE industry for over two decades, as documented by the Open Web Application Security Project (OWASP)[42] [139]. Even with various defence mechanisms, new injection threats continue to emerge [139]. Conversely, security violations are rare in Java code snippets, aligning with the study by Meldrum et al. [11] where Java security vulnerabilities are somewhat minimal. In terms of total violation counts, *SELECT ** was flagged most frequently in SQL, which risks accidental exposure of more data than intended [140]. This suggests a trade-off between efficiency and security, where users might prioritise brevity over explicit column names [118]. Similarly, JavaScript exhibits frequent object injection hotspots due to a lack of input validation, opening doors for cross-site scripting (XSS) attacks [141]. Python shows misuse of the *random*[43] standard library (*S311*), with developers potentially unaware that it generates pseudo-random numbers instead of truly random ones [142]. This could have security implications if the generator's state is compromised (known as state compromise extension attacks) [143]. Unsanitised URL inputs are common in Ruby, raising the risk of injection attacks, unauthorised access (e.g., root access) and telnet daemons [144]. Finally, Java developers tend to omit the final modifier for public static fields (*MS_SHOULD_BE_FINAL*). While the contents of the object can change, the reference itself cannot be reassigned, limiting unintended access reference [11]. This aligns with Meldrum et al. [11], who found mutable static fields being a prevalent security concern in Java. Overall, the majority of security violations stemmed from unsanitised inputs, which can manifest in various ways depending on the programming language. For instance, SQL code might be vulnerable to string concatenation attacks, while JavaScript could be compromised through techniques like non-literal requires. Similarly, Ruby code might be susceptible to security risks if it employs the *Eval* function. Another category of security violations involved hardcoded passwords within the code itself. Examples include readable passwords found in SQL snippets, or the presence of hardcoded credentials (e.g., *S105* in Python and *DMI_CONSTANT_DB_PASSWORD* in Java). We also found certain threats that are more severe albeit less frequent: violations related to bidirectional characters in JavaScript code, which can evade human detection during code review [145].

[42] https://owasp.org

## 5.2 Code Snippet Violation Densities (RQ2)

In this section, we discuss snippet violation densities as outlined in RQ2: *How do coding practices among US contributors differ across states and cities in the United States?*

Looking at our results, the size of the tech industry and economy seems to have an effect towards the quantity of parsable code snippets. States like New York, California, and Texas, renowned for their sizable tech workforces (e.g., Texas with a 5.5% net growth in tech jobs in 2022) [146], yielded a higher number of parsable snippets, which can likely be attributed to the volume of code production in these regions. Interestingly, these same states (New York, California, and Texas) exhibited lower violation rates across all code quality dimensions and programming languages. Particularly for reliability and readability, their violation densities tended to cluster around the median, suggesting a balance between non-compliance (i.e., high violation density) and adherence to coding best practices (low violation density). One possible explanation for this observation lies in the age and experience diversity of the tech hubs' coding communities, which ranges from seasoned professionals to apprentices [147]. While senior developers are more likely to adhere to established practices, resulting in reliable and readable code, contributions from less experienced individuals might introduce deviations from these standards [148, 149]. **Therefore, we posit that this age diversity within the workforce might influence the overall conformance (or departure) from best practices**. California and New York also exhibit lower violation rates in performance and security dimensions, which may be attributed to the fiercely competitive nature of these tech ecosystems. Being home to giants like Google, IBM, and Bloomberg might incentivise a stronger emphasis on performance and security best practices [150, 151], which could be seen as a way to gain a competitive advantage in the software market and attract consumers who prioritise a smooth and secure user experience [177]. Moreover, as tech hubs are often at the forefront of cutting-edge technology [152, 153], projects in such regions might prioritise performant and secure code to remain competitive. For instance, enhanced code security was found to mitigate data leaks, contributing to user retention [150].

Conversely, states with less established tech sectors, like Vermont or North Dakota, showed higher readability violation densities. This might be because smaller companies lack the formalised coding conventions found in larger firms (e.g., Google Java Style[44]), leading to more stylistic inconsistencies. Interestingly, **states without a tech presence exhibited lower reliability violation rates compared to**

[43] https://docs.python.org/3/library/random.html

[44] https://google.github.io/styleguide/javaguide.html

**their tech-heavy counterparts**. One potential explanation is their prioritisation of code functionality rather than business-related factors (e.g., time-to-market) that are otherwise prevalent in startup-intensive environments [154]. Finally, our findings hinted at a trend beyond just the presence of tech hubs. States like Maryland and Washington, with varying tech industry sizes, displayed the lowest performance violation rates for JavaScript and Ruby. A closer investigation revealed that these states offer some of the highest salaries for web developers [155]. For instance, Washington boasts the highest median web developer salary nationwide ($138,780), followed closely by Virginia ($101,060) [155]. Considering the prevalence of JavaScript and Ruby in web development [156], this observation hints at a potential link between developer compensation and adherence to best practices. In particular, developers may be incentivised to enhance their JavaScript and Ruby performant coding practices to remain competitive in a high-paying job market.

Our city-level analysis revealed similar, though less pronounced, patterns. Tech hubs with large R&D clusters (e.g., San Francisco, CA and Seattle, WA) displayed moderate levels of reliability and readability violations. This aligns with our prior findings where diverse workforces in these regions, with both experienced and junior developers [147], influence how violations are flagged [148, 149]. Interestingly, **large cities (not necessarily dominated by tech), exhibited higher performance violation rates than smaller ones**. This contrasts with findings at the state-level, which could be due to the specific industries prevalent within these cities. Unlike Silicon Valley with its deep tech roots, cities like Irving, TX, whose largest employers reside in sectors like investment banking (Citigroup; 6,162 total employees) and insurance (Allstate Insurance; 3,068 total employees) [157], may not possess a strong tech identity. These non-tech industries may prioritise functionality over speed, which may translate to higher performance violations for users in these locations. Security violations, however, mirrored readability and reliability findings. Larger cities with a strong R&D presence, like Houston, TX exhibited lower security violations. This emphasis on security likely stems from the inherent risks associated with data during the rapid experimentation that characterises R&D activities [150, 158].

### 5.3 Code Snippet Relations (RQ3)

In this section, we discuss how diversity indicators affect violation densities as outlined in RQ3: *How do diversity indicators of US regions affect code quality?*

Our findings suggest that certain diversity aspects indeed affect code quality. One key finding emerged from examining socioeconomic aspects, as measured by the indicator *Gini Index* which reflects regional wealth distribution [159]. We found a negative correlation where regions with more equitable wealth distribution had lower violation densities. This correlation could be partially explained by increased investment in education within these regions, which aligns with Wilterdink [160] where it is argued that decreased expenditure on public education may inherently speed up wealth inequality. Inversely, **regions with more equitable wealth distribution might allocate more resources towards education, fostering workforces with the skills necessary to write overall better code with fewer errors**. For instance, Florida's substantial investment in higher education (approximately 5.691 billion USD in 2021) – one of the highest figures nationwide [161] – also exhibits a comparatively high *Gini index* of 0.43 (3rd highest across all states) and small violation densities for Ruby Readability at 1.118. Another interesting finding regarding socioeconomic diversity is the possible association between higher regional income (measured by *per capita income in the past 12 months*) and more secure coding practices, particularly evident in Java and Python snippets. Wealthier regions might have an advantage in attracting developers with strong cybersecurity expertise, or given the high cost of implementing proper cybersecurity practices [162], they may be able to more effectively invest in training programs focussed on secure coding practices. **Thus, we postulate that firms residing in these high-income regions are better-equipped to commit these investments**. Moreover, the paramount importance of cybersecurity in enterprise applications (which are often developed using these popular languages [163, 164]) corroborates our findings.

Next, we found a positive correlation between ICT domestic access and code quality, evident within the indicators *households with one or more types of computing devices* and *Internet subscriptions in household: with an Internet subscription*. Looking closer, it appears that this correlation holds regardless of language and dimension, which suggests that any potential causal influence is broad. In fact, our findings somewhat align with Silva et al. [165] who suggest that early domestic access to ICT, such as computers in childhood, fosters the development of computational thinking skills in later life. In the context of our study, we postulate that **problem-solving and algorithmic pedagogic approaches that give way to programming and robotics [165] would eventually lead to the production of higher quality code**, as initial findings suggest. We also found a negative association between the percentage of white population in a state and the readability of Java snippets. However, this observation must be considered with extreme caution, as it is highly unlikely that this correlation is solely attributable to racial distribution. Exploring potential mediating factors instead sheds light into more subtle socioeconomic facets. For instance, higher percentages of white population are mostly found in more rural states (e.g., Vermont at 94.89%, West Virginia at 93.01%) [12], and these rural states suffer from lower investment in infrastructure and education, as well as high income disparity [166]. In the context of our results, **limited educational coverage may explain the observed prevalence of readability violations in Java code, which are otherwise found in more urban states** such as Nevada (81.50% white population) and California (72.46%). Similarly, our findings regarding how educational attainment affect code quality extends to the indicator *school enrolment by level of school for the population 3 years and over: enrolled up to grade 12*. Substantiating the prior observation regarding limited educational resources in rural areas, this indicator also revealed that higher completion rates of compulsory basic education is associated with more performant snippets. We can therefore infer **that a well-educated population may contribute to a stronger foundation in coding practices**. In fact, several developed nations including Singapore and South Korea, have already begun integrating coding into their K-12 curriculum [167]. This decision is rooted in the rationale where introducing such concepts from an early age will benefit even those whose aim is not to be professional programmers [167], as these skillsets foster creativity and advance their spatial and reasoning skills [168].

In terms of gender, we found states with a higher male proportion tend to also have fewer reliability violations regardless of language. **Due to the perceived reliability of code snippets, this could potentially explain the observation that males tend to accrue more reputation points**, as outlined by Vasilescu et al. [169]. In contrast, we also found states with higher female proportions tend to also yield fewer readability violations. Some studies propose that women may be more likely to exhibit traits associated with meticulousness and attention to detail [170, 171]. **We argue this inherent quality could translate into a collective tendency towards writing more readable snippets.** However, deeper investigation might be needed to ascertain the nature of these differences. **Furthermore, we observed that states with a higher proportion of females tend to output more performant code.** This observation is notable considering that code performance is often neglected in the face of more tangible factors such as usability and design patterns [172, 173, 174]. Our results contribute to the existing body of knowledge by adding another perspective that neglecting performance is not as widespread as prior works previously hinted. Finally, previous research indicates that both genders can indeed contribute to more secure coding practices, albeit through different approaches [175]. For example, males may favour techniques like obfuscation, cryptography, and secure privilege management, while females might lean towards secure networking, library utilisation, and data minimisation practices [175]. This is also reflected in our results, where the gender composition of a state appears to have no noticeable impact on its level of security violations, unlike the other three dimensions. That is, our findings suggest that states with higher female composition exhibit similar levels of security violations as those with higher male composition.

## 5.4 Code Snippet Themes (RQ4)

In this section, we discuss the latent themes inherent in each regions' code snippets as outlined in RQ4: *What themes are prevalent in code quality violations, and how do these themes differ across US regions?* Looking at users' code snippets across regions, we found that users of California tend to contribute snippets that lean towards complexity and implicitness, reducing intuitiveness for new programmers in the corresponding languages. The majority of their reliability errors seem to align with this observed complexity, such as unclosed resources and race conditions. One potential explanation for this finding could be the established presence of major tech hubs and a thriving tech ecosystem within California. We postulate that **developers of such environments might be more likely to produce code that reflects professional practices, which may differ from beginner-oriented approaches**. In fact, California held the highest number of tech jobs (1,487,864) in 2023, almost twice that of the second-highest, Texas (867,278) [146]. Our findings add another nuance to prior RQ2 results. Previously, we found that states with broader tech ecosystems (e.g., California and New York) exhibited moderate densities of reliability and readability violations. Yet a closer examination of the actual code snippets from these states reveals a trend towards more complex violations, contrasting the lower overall frequency. **These findings thus indicate that developers in these areas may prioritise professional coding practices that can lead to more intricate and less explicit, yet less error-prone code**. This pattern corroborates Bowen et al. [176], who argued that enterprise software systems need to be reliable. However, snippets from California were found to exhibit slight performance overheads or employed less efficient methods compared to faster alternatives. This trend might be linked to the fast-paced startup culture prevalent in California where developers are expected to deploy software products rapidly [177]. We posit that this behaviour translates to Stack Overflow where **users in such environments may prioritise functionality over minor performance optimisations, as long as these overheads are not detrimental to the software's functionality**. Additionally, the pressure to deliver software quickly might lead to the adoption of less efficient but readily implementable solutions, as opposed to investing time in complex yet marginally faster alternatives [178]. Certain specific violations were also notable. For instance, code snippets from California were found to yield high degrees of redundant stylistic choices, which might again be attributed to the state's large number of tech companies (55,868 in 2023) [146], inadvertently leading to a diversity of coding conventions. In such dynamic environments, it may be unlikely that all developers will strictly adhere to a single coding standard, leading to the stylistic deviations as observed in our results.

User contributions from Utah exhibited the lowest frequency of performance-related violations. This finding might be explained by the state's industrial landscape, which is centred around performance-critical sectors like mining, manufacturing, and petroleum production, alongside information technology. **The prominence of these industries, exemplified by the Bingham Canyon Mine which is one of the world's largest [179], could foster a culture of prioritising code efficiency** within the state's developer community. Certain initiatives also exist to bridge the gap between IT and manufacturing practices, such as the century-old Utah Manufacturers Association[45]. At the other end of the spectrum, user contributions from North Dakota primarily exhibited simpler reliability violations, such as syntax errors and generic raw exception handling. This observation aligns with the state's nascent tech ecosystem as it only hosted 1,219 tech business establishments and 13,272 net tech employment in 2023 [146]. However, the presence of more complex violations, like misusing built-in methods, suggests a growing sophistication within North Dakota's developer community and hints that this state could catch up with others. In fact, in 2022, it experienced a net growth of +2.5% in tech employment, which is relatively higher than other less rural states such as Massachusetts (+1.6%) and District of Columbia (+0.6%) [146]. Interestingly, our analysis revealed a similar tendency for users in both California and North Dakota to contribute code with inadvertent file openings via insecure connections. This aspect highlights another key finding of our qualitative analyses where, **despite the vast differences in the technological landscapes between states, developers seem to exhibit a degree of 'hive mind,' such that they tend to have similar thought patterns** regardless of where they reside. These results align with Salminen et al. [180] who outlined the uniformity in software developer thinking, suggesting knowledge transfer or common pitfalls regardless of where users reside. Other coding themes were also found to exhibit consistent patterns across all three states. One example is the low prevalence of storing passwords in plaintext or unsalted hashes, which hints that users across all three states may already be aware regarding the dangers of such practices.

## 5.5 Implications

Our study holds significance for both software development teams, Stack Overflow administrators, and other CQA platforms that foster collaboration. Firstly, different programming languages have varying inherent language features that may influence the types of violations users make. For example, Java snippets exhibited a higher prevalence of reliability errors, potentially due to the language's stricter syntax compared to Ruby, which showed a bias towards readability errors given its more flexible conventions. Software development teams should consider the inherent complexity of their chosen languages when building their tech stack. Additionally, implementing standardised coding rulesets can help mitigate errors stemming from these language-specific characteristics. For example, off-the-shelf tools such as Codacy, Semgrep, and SonarQube provide extensible rulesets such that teams can add their own standards on top of existing ones. Next, regions with more established tech industries tend to exhibit more intricate coding practices, leading to more complex errors. We underscore the importance of implementing effective onboarding mechanisms for geographically dispersed software development teams. Such mechanisms can help bridge potential knowledge gaps between team members from different regions, ensuring everyone has a solid understanding of the team's coding conventions, best practices, and what constitutes good quality code.

[45] https://www.linkedin.com/company/utah-manufacturers-association

Furthermore, there is also a need for targeted educational initiatives in regions with less established tech sectors, such as coding-related workshops. We believe such programs could address the prevalence of simpler errors, and in turn equip these regions' tech workforce with stronger programming fundamentals. Policy makers can allocate more investment in education as our results hinted that such endeavours may improve code quality in rural states. In terms of gender, our results revealed contrasting coding styles. For instance, females were associated with more readable code, whilst males were associated with higher reliability. Software development teams can leverage these insights to integrate diverse strengths and perspectives within the codebase, ultimately leading to a more well-rounded and robust product. For Stack Overflow (and similar CQA) site administrators, we advocate for a built-in linting feature that automatically assesses code snippets for reliability, readability, performance, and security issues before users submit them in their answers. We believe such a feature could promote higher-quality code examples within the platform and reduce inadvertent errors for users who integrate these snippets into their codebases.

# 6 THREATS TO VALIDITY

This section outlines the limitations to our work. We classify threats into internal, external, and construct validity, in line with prior literature [23].

## 6.1 Internal Validity

Firstly, 14,174,843 users were dropped from the dataset due to missing or invalid location data, and another 2,986,774 were excluded due to their location data pointing to other countries. Among these excluded users, some may have been from the United States, even if not explicitly stated. We acknowledge that these actions taken to ensure representativeness of our analyses may inadvertently omit valuable information. Conversely, there is also the possibility of users entering false locations. Secondly, while our work is influenced by Meldrum et al. [11] as we used roughly similar quality dimensions and linting tools, we omit the calculation of specific data attributes for each snippet other than LOC and LLOC (e.g., code length, code spaces, or snippet-per-answer ratio), as was done in their study. The rationale behind this is that programming languages vary in their complexity, and therefore such attributes may not be comparable. For instance, a "Hello World" program would be more concise in JavaScript than in Java. Third, our calculation of LLOC for each snippet incorporated regular expressions to comprehensively capture multiline comments, mindful of the variation in comment identifiers across different languages. While we acknowledge the potential for some comments to evade detection, we have crafted and rigorously tested our regular expressions to minimise this risk to the greatest extent possible. This approach ensures a high degree of confidence in the overall accuracy of our LLOC measurements. We also acknowledge potential reporting discrepancies with respect to the Ruby Readability dimension, with the top ten most frequent violations encompassing only 55.99% of the total, while other languages and dimensions exceeded 75% coverage. Thus, we might have missed some important details regarding how Ruby readability varies across different regions. Lastly, some tools employed priority indices such as PMD, while other tools did not. We postulate that including these indices in the calculation process may introduce an element of unfairness, even when employing weighting mechanisms. Consequently, we opted to exclude such indices from our results and treat all violations across all languages as having equal priority.

## 6.2 External Validity

Our study is focussed only on Stack Overflow users inside the US, excluding users from other countries. Our findings may therefore not be generalisable towards users from countries outside the US, or towards other CQA sites like Cross Validated or Ask Ubuntu. Secondly, Stack Overflow is a dynamic platform where coding practices may evolve over time due to changes in technology trends and user demographics. To ensure results align with the latest trends, we have used the June 2022 release of the Stack Exchange Data Dump, being the latest at the time of our investigation. While our findings may not accurately capture the current state of software development in the US, prior work has noted that Stack Overflow data are consistent across time [11, 181]. Other threats still loom, however, such as the evolution of code quality due to the emergence of generative models like Google's Gemini or OpenAI's ChatGPT [26]. Over half of GPT-generated answers contain errors [26], underlining Stack Overflow's continued relevance as a source of reliable solutions. In fact, both of our quantitative and qualitative explorations reveal that reliability errors are primarily program flow alterations, not critical issues that prevent the code from functioning. Therefore, these errors can likely be addressed when users integrate the snippets into their projects. Thirdly, our selection of five languages poses a limitation to generalisability by excluding other popular languages like C++ or C#, and emerging languages like Rust or Go. Inferences drawn from our analyses may not be applicable to these other languages. For instance, TypeScript snippets might not necessarily exhibit the same security violations as JavaScript snippets, despite the syntactic similarity of both languages. Our language selection also only includes two of the top ten most popular languages in the 2022 Stack Overflow Developer Survey [81], namely Python and SQL. Fourth, across all languages, we observe a substantial degree of unparsable and uncompilable snippets, impeding our ability to assess them. Consequently, our results may not be a representative depiction of these five languages. However, the identified trends largely align with prior literature [11, 122], so we believe our work still provides sufficient insights for SE research and practice. Our fifth threat to external validity pertains to the small size of Stack Overflow snippets. These snippets are often provided in isolation, lacking full context of the software projects they belong to (if any). While literature has established snippets' presence in large software repositories [15, 16], inferences from this study might not extend beyond the specific context of our work. Finally, our study samples California, Utah, and North Dakota to conduct a series of inductive content analyses, mirroring prior research [12, 13]. This selection serves as a representative proxy due to the impracticality of analysing content from all states. Consequently, our findings may not be universally applicable to other regions, and there is a possibility of encountering distinct regional trends if content analysis were to be conducted on other states.

## 6.3 Construct Validity

To confirm whether concepts are operationalised sufficiently [182], we initially evaluated the accuracy of Guesslang, and our assessment indicates that the tool performs satisfactorily (as described in Section 3.3). This evaluation aimed to reduce the risk of mislabelling languages with near-identical syntax (e.g., distinguishing between C/C++, or JavaScript/TypeScript/CoffeeScript). Despite our efforts, some threats persist, as illustrated by the embedded SQL snippet shown in Figure 5. Afterwards, we noted that linting tools were able to identify unparsable snippets, mitigating this particular threat to a minimal level. The classification of the 54 technologies predicted by Guesslang may also be subject to errors. For instance, some developers may not consider SQL to be a programming language as it cannot be used to create standalone applications. However, by aligning our categorisations with existing literature, we ensure that this threat is minimal.

With regards to our quality dimensions, we acknowledge that other researchers and practitioners may have different interpretations for these constructs. In other words, what constitutes reliable, readable, performant, or secure code may vary from person to person. While our four quality dimensions were formulated based on prior literature, it remains probable that these four facets may not fully capture the complex nuance of code snippet quality. For instance, there may exist additional dimensions that were not accounted for in our study. After all, measuring code quality remains a challenging endeavour [63].

In terms of our inductive content analyses, we achieved inter-coder reliability with $\kappa$ values between 0.825 and 0.869, signifying a good level of agreement [115]. While the level of agreement is strong [183], caution is warranted when interpreting these results as the values fall into the range of 64–81% reliability [183]. In other words, 19–36% of coded exchanges may not be as reliable [183]. Additionally, the final coding themes represent a refinement process where some potential themes may not have been included. However, all disagreements between coders were addressed through consensus to ensure the final coding scheme's robustness, and we believe this iterative process minimises this threat.

# 7 CONCLUSIONS AND FUTURE WORK

This study examines code snippet quality within Stack Overflow answers, focussing on SQL, JavaScript, Python, Ruby, and Java. While prior research has examined related concepts, most of these studies have focussed on only one or two programming languages, and have typically assessed only one quality dimension. Our study aims to advance the conceptualisation of code snippet quality by encompassing a broader spectrum of programming languages and evaluating it across four distinct quality dimensions: reliability, readability, performance, and security. Moreover, we also investigate how code quality may relate to diversity complications that are widespread in SE-related practices. We found evidence that readability violations are the most prevalent across all languages, followed by reliability, performance, and security. While readability violations largely pertain to issues like code clutter and improper whitespace, more nuanced violations also exist, such as excessively long code lines that induce confusion due to prolonged lateral eye movements. In terms of reliability, violations typically manifest either as critical errors leading to execution interruptions, or as subtle alterations in program behaviour. Performance-related violations vary in their severity, ranging from small memory allocations to catastrophic backtracking when constructing regular expressions [138]. Finally, security violations mainly revolve around unsanitised string inputs (posing risks of code injection and XSS attacks), along with less common but still severe violations such as Unicode bidirectional attacks (leading to trojans) and mutable static fields.

We calculated violation densities for each dimension to ensure fair representation when analysing results across states and cities of the US. The US was selected as a case study given its role as the birthplace of Stack Overflow, its status as the largest English-speaking country (with English being the *lingua franca* of SE), its global leadership in technology, and its position as host to the highest number of Stack Overflow users. Firstly, states with a large tech presence (e.g., California, Texas, and New York) tend to generate more parsable snippets, which may be attributed to the size of their tech workforce, translating to higher code production and increased presence on Stack Overflow. Contrary to expectations, the abundance of parsable snippets does not directly lead to a higher incidence of violations across different languages and dimensions. Notably, such states displayed only moderate violation densities in reliability and readability dimensions, and even less for performance and security. We propose such variability stems from the combined influence of a diverse tech workforce (i.e., senior and junior programmers), emerging industries, and the ubiquitous nature of technology within these regions. Users from rural areas exhibit a lower propensity for generating reliability violations, which may arise from the prioritisation of reliable functionality over rapid release cycles (time-to-market), with the latter often characteristic of startup environments in high-tech states. Our analysis reveals a predominantly top-down trend propagation, where state-level patterns are largely reflected at the city level (albeit not as stark). Cities renowned for their R&D clusters such as San Francisco, CA and Seattle, WA generated moderate levels of reliability and readability violations, and even fewer under security. The trend diverges for performance violations, however, with large, non-tech-focussed cities having more – which could be due to the specific industries in those areas. In terms of diversity indicators, regions with more equitable wealth distribution, higher education investment, and higher coverage of domestic technology exhibited fewer code violations. Interestingly, the racial makeup of a state itself was not a strong indicator, but rather the socioeconomic factors often associated with predominantly white populations, such as lower investment in rural areas. Gender also played a role: states with a higher proportion of females had fewer readability and performance violations, while those with a higher proportion of males had fewer reliability violations. Security showed an interesting trend where both genders contributed to better code, but through different approaches. Qualitative content analyses confirmed that developers from California, a state with a large tech hub, tended to write more complex code and prioritised professional practices over readability for beginners. This leads to errors that are also equally complex. Their fast-paced startup culture that emphasises time-to-market over optimisation also gives rise to minor performance inefficiencies (albeit not detrimental). In contrast, Utah, which hosts performance-critical industries, had users who contributed code with a higher emphasis on efficiency. North Dakota, whose tech industry is nascent, exhibited simpler errors but also certain intermediate-level ones, suggesting a maturing developer base. Interestingly, some coding patterns were consistent across all regions, such as the avoidance of storing passwords in plaintext or unsalted hashes, highlighting a shared knowledge base among developers regardless of location. Nevertheless, our findings suggest that users contributing answers on Stack Overflow may not always fully grasp the broader reliability, readability, performance, and security implications of their given solutions. Thus, we advocate developers to exercise caution when integrating such code into their software repositories. Although the majority of identified violations are not of critical concern, their unchecked presence has the potential to adversely impact the repository's overall quality.

Our study has illuminated potential directions for future studies. Subsequent investigations could, for example, leverage the insights gained from our study to employ Bayesian inference techniques to complement our frequentist approaches. It is also possible to triangulate results from all four dimensions to determine whether there is a trade-off between each quality dimension. For instance, researchers may validate whether there is any causal relationship between performance and readability, or whether performance and security are inversely proportional [52]. Notably, an additional avenue of research could involve the development of a composite measure that integrates all four operationalised constructs, resulting in an all-encompassing metric that holistically conveys the idea of code snippet quality. There is also the need to conduct a full-scaled qualitative investigation to ascertain the unearthed patterns from our content analysis. Furthermore, leveraging our facets of quality, future research could employ modelling techniques to quantify the extent of user contribution within the platform. We contend these research venues hold significant value in enhancing our understanding of the coding practices prevalent on the platform, thereby elucidating their implications for broader software development practices. Finally, we assert that our findings should not, under any circumstances, be utilised to profile the superiority of one region over another.